\documentclass[reprint, twocolumn, aps, prd, eqsecnum, amsmath, amssymb, amsfonts, superscriptaddress, nofootinbib, preprintnumbers,
]{revtex4-2}

\usepackage[T1]{fontenc}
\usepackage{hypernat}
\usepackage[colorlinks=true, allcolors=teal]{hyperref}
\usepackage{physics}
\usepackage{tikz}
\usepackage{ragged2e}
\usepackage{subcaption}
\usepackage{dsfont}
\usepackage{mathtools}
\usepackage{orcidlink}
\usepackage[normalem]{ulem}

\newcommand{\mc}{\mathcal}

\begin{document}

\title{R\'enyi entropy of single-character CFTs on the torus}

\author{\firstname{Luis Alberto} \surname{Le\'on Andonayre}\,\orcidlink{0000-0003-4912-5675}}
\email{lal10@hi.is}
\affiliation{Science Institute, University of Iceland, Dunhaga 3, 107 Reykjav\'ik, Iceland.}

\author{\firstname{Rahul} \surname{Poddar}\,\orcidlink{0000-0001-7602-455X}}
\email{rahul.poddar@unf.edu}
\affiliation{Department of Physics, University of North Florida, Jacksonville, Florida 32224, USA.}

\preprint{arXiv:2412.00192}
\date{\today}

\begin{abstract}
We introduce a nonperturbative approach to calculate the R\'enyi entropy of a single interval on the torus for single-character (meromorphic) conformal field theories. 
Our prescription uses the Wro\'nskian method of Mathur, Mukhi, and Sen \cite{Mathur:1988rx}, in which we construct differential equations for torus conformal blocks of the twist two-point function. 
As an illustrative example, we provide a detailed calculation of the second R\'enyi entropy for the $\rm E_{8,1}$ Wess-Zumino-Witten (WZW) model.
We find that the $\mathbb Z_2$ cyclic orbifold of a meromorphic conformal field theory (CFT) results in a four-character CFT which realizes the toric code modular tensor category.
The $\mathbb Z_2$ cyclic orbifold of the $\rm E_{8,1}$ WZW model, however, yields a three-character CFT since two of the characters coincide.
We then compute the torus conformal blocks and find that the twist two-point function, and therefore the R\'enyi entropy, is two-periodic along each cycle of the torus.
The second R\'enyi entropy for a single interval of the $\rm E_{8,1}$ WZW model has the universal logarithmic divergent behavior in the decompactification limit of the torus, as expected as well as the interval approaches the size of the cycle of the torus.
Furthermore, we see that the $q$-expansion is UV finite, apart from the leading universal logarithmic divergence. 
We also find that there is a divergence as the size of the entangling interval approaches the cycle of the torus, suggesting that gluing two tori along an interval the size of a cycle is a singular limit.
\end{abstract}

\maketitle

\section{Introduction}
\label{sec:introduction}

Quantum entanglement is a fundamental property of quantum systems \cite{Reeh:1961ujh}, with significant relevance across various areas of physics, including condensed matter physics \cite{Osborne:2002zz, Vidal:2002rm, Latorre:2004pk, Plenio:2004he, Cramer:2005mx}, quantum information theory \cite{Wolf:2007tdq}, black holes physics \cite{Bombelli:1986rw, Srednicki:1993im, Callan:1994py, Almheiri:2020cfm}, quantum field theory (QFT) \cite{Calabrese:2004eu, Casini:2004bw, Kitaev:2005dm, Casini:2017vbe}, and holography \cite{Ryu:2006ef}.
Over the past 20 years, entanglement has become a crucial area of study, particularly through quantum informational measures such as entanglement entropy \cite{Ryu:2006bv, Calabrese:2004eu, Ryu:2006ef, Kitaev:2005dm}. 

Since the discovery that the entropy of a black hole is proportional to their area \cite{Bekenstein:1973ur}, understanding how entropy and entanglement arise in systems has been crucial. 
This shows that black hole microstates are associated with the event horizon and should help us to gain some insight into the Hilbert space of quantum gravity \cite{Strominger:1996sh}.
However, it is still a challenge to compute measures of entropy and entanglement in interacting systems and systems on higher-genus Riemann surfaces.

For QFT, some measures of quantum entanglement are not well-defined, as there is an infinite amount of quantum entanglement between any two given subregions.
This can be quantified by noting that the algebra of local operators for a quantum field theory without a UV cutoff has a von Neumann factor of type III$_1$ \cite{Guenin1963, Bisognano:1975ih, Bisognano:1976za}. 
In other words, without imposing a UV cutoff, objects like reduced density matrices and traces over subregions cannot be defined in a quantum field theory, and the Hilbert space of a QFT cannot be factorized into separate Hilbert spaces for each spatial subregion.  
For two-dimensional conformal field theories (CFTs) on the plane, it is well known that the entanglement entropy for single intervals satisfies a universal UV divergent behaviour \cite{Cardy:1988tk}. 
Computing measures like entanglement entropy becomes much harder on higher genus surfaces and with a larger number of intervals for interacting systems.

Entanglement entropy, or von Neumann entropy, for a subregion $A$ is defined by
\begin{equation}
  S_{\rm vN}(A) = -\Tr(\rho_A \ln |\rho_A|), 
\end{equation}
where $\rho_A$ is the reduced density matrix for the subregion $A$.
In practice, this quantity is difficult to compute.
A computationally useful measure of quantum information is the R\'enyi entropy \cite{renyi2007probability}.
To compute this, we employ the replica trick, which involves replicating the field theory and sewing together the surfaces by identifying the ends of the interval of interest. 
This can be achieved by introducing a branch cut between the ends of the intervals, where the replicated theory is defined on the corresponding Riemann surface. 
The $N$th R\'enyi entropy,
\begin{equation}
  S_N(A) = \frac{1}{1-N} \ln\left| \Tr \rho_A^N\right| = \frac1{1-N} \ln\left| \tfrac{Z(A,N)}{Z(0,N)}\right|,
\end{equation}
is in terms of the reduced density matrix $\rho_A$ for the subregion $A$, when possible to define (i.e., with a UV cutoff or for operator algebras of type I or II), or in terms of the partition function on the Riemann surface of the replicated theory $Z(A,N)$.
The normalization $Z(0,N)$ is the replicated surface without sewing together the replicated copies. 
On the plane, $Z(0,N) = Z^N$, where $Z$ is the partition function of the unreplicated theory.
The entanglement entropy, or von Neumann entropy, can be recovered with an analytical continuation in the limit,
\begin{equation}
  \lim_{N\to1} S_N(A) =  S_{\rm vN}(A). 
\end{equation}

To work with the replicated theory, one must compute the $\mathbb Z_N$ cyclic orbifold of the original theory.
For a CFT, this introduces new primary fields, corresponding to the new twisted sectors, or heighest weight representations of the new $\mathbb Z_N$ symmetry.
These are known as twist operators $\sigma$ \cite{Atick:1987kd, Klemm:1990df}.
Inserting a twist operator introduces a monodromy $e^{2\pi i/ N}$ at the insertion point. 
Therefore, when inserting a twist-antitwist pair, this introduces a branch cut between the insertion points. 
One can then work out correlation functions of these twist operators, to compute the partition function of the replicated CFT on the Riemann surface created by introducing the branch cut \cite{Calabrese:2004eu},
\begin{equation}
  Z(A,N) = \ev{\sigma(z_1,\bar z_1) \overline \sigma(z_2, \bar z_2)}.
\end{equation}
Here, the subregion $A$ has end points $(z_1,\bar z_1)$ and $(z_2,\bar z_2)$ on the unreplicated Riemann surface. 

On the plane, we can use conformal invariance to compute this two-point function trivially, which allows one to make a universal statement about the R\'enyi and von Neumann entropies for conformal field theories on the plane. 
This universal behavior is no longer true for theories on higher-genus surfaces. 
R\'enyi entropies on the torus have only been calculated exactly for free bosons and free fermions, and have been perturbatively calculated for other interacting systems. 
Holographic techniques have also been used to calculate the entropy of CFTs at large central charge \cite{Ryu:2006bv, Headrick:2010zt, Barrella:2013wja, Datta:2013hba, Haehl:2014yla}.
Torus correlators in the holographic or semiclassical limit have also been computed \cite{Kraus:2017ezw, Alkalaev:2017bzx, Alkalaev:2018qaz, RamosCabezas:2020mew, Alkalaev:2020yvq, Alkalaev:2022kal, Pavlov:2023asi,Alkalaev:2023evp}. 

To compute the entanglement entropy of disjoint intervals on a torus for free bosonic and fermionic theories, strategies involving explicit computations of propagators and Green's functions have been employed. 
For a CFT of free bosons, it is possible to explicitly compute Green's functions by means of cut-differentials and Ward identities on the torus.
The problem is broken down into computations of the classical and quantum part, respectively, and has been tackled accordingly \cite{Atick:1987kd, Datta:2013hba, Chen:2015cna}.
Corrections to the entanglement entropy on the torus for $(d+1)$-dimensional generalized quantum Lifshitz models have been calculated in \cite{Angel-Ramelli:2019nji}.

Similarly, the task of computing the entanglement entropy of free fermions on the torus has been done by considering the different boundary conditions of the twisted sectors and computing propagators in these twisted sectors.
Alternative approaches have employed resolvent analysis, enabling the determination of modular\footnote{The term ``modular'' here refers to the modular operators of Tomita-Takesaki theory \cite{Haag:1996hvx}.} data associated with a local subregion \cite{Fries:2019ozf, Erdmenger:2020nop,Fries:2019acy, Blanco:2017akw, Lashkari:2015dia, Sarosi:2017rsq}.
Constraints from modular invariance on the entropy of fermions on the torus were studied in \cite{Lokhande:2015zma}.
Furthermore, interesting results relating the Jacobi and the Siegel theta functions have arisen from computing the R\'enyi entropy of fermions on the torus \cite{Mukhi:2017rex, Mukhi:2018qub}.
Interesting extensions include studying nonrelativistic free fermions on the torus \cite{Aguilera-Damia:2023jyc}.

However, the methods listed above are useful only for free theories, or in the semiclassical limit. 
In this paper, we employ a method to construct a differential equation for the conformal blocks of a correlator on the torus, introduced by Mathur, Mukhi, and Sen \cite{Mathur:1988rx}.
This is tractable for CFT correlators with a small number of conformal blocks since the order of the differential equation constructed is equal to the number of linearly independent conformal blocks.
The advantage is that this method is constrained by modular invariance and double periodicity on the torus, which allows one to explicitly construct the differential equation, which may be solved in a variety of techniques. 
This procedure has also spawned a fruitful classification program of rational conformal field theories (RCFTs), where the differential equation constructed is a ``modular linear differential equation'' or MLDE for short \cite{Mathur:1988na, Mathur:1988gt, Naculich:1988xv, Hampapura:2015cea, Gaberdiel:2016zke, Hampapura:2016mmz, Chandra:2018pjq, Chandra:2018ezv, Mukhi:2020gnj, Mukhi:2022bte, Das:2022uoe, Das:2023qns}.
MLDEs are highly constrained by modular invariance, whose solutions are characters of potential RCFTs. 
Recently, RCFTs have attracted significant attention beyond the high-energy physics community, particularly in the context of the Chern-Simons/Wess-Zumino-Witten duality applications related to the fractional quantum Hall effect and in the implementation of fault-tolerant quantum computing with non-Abelian anyons \cite{RevModPhys.80.1083, Dong:2008ft, PhysRevB.100.085116}.

To determine the twist two-point function, we focus on CFTs, which yield a low number of conformal blocks. 
Therefore, as a first step in using this method to compute R\'enyi entropies, we concentrate on single-character, or meromorphic, CFTs. 
A meromorphic CFT has a single primary under the extended current algebra, which is the vacuum, and therefore a single character corresponding to the vacuum primary \cite{Goddard:1989dp}.
Since the partition function must be modular invariant, the single character also must be modular invariant, up to a phase.
Recall that characters are defined by
\begin{equation}
  \chi_i(\tau) = \Tr_{{\mc H}_i} \left(q^{L_0 - \frac{c}{24}}\right),
\end{equation}
where $q=e^{2\pi i \tau}$ is the nome, and the trace is over the Verma module corresponding to the primary field. 
Under the modular $\mc T$ transformation, which sends $\tau \to \tau + 1$, the characters have an eigenvalue of $e^{2\pi i (h_i - \frac c{24})}$, where $h_i$ is the conformal dimension of the primary.
The vacuum has a conformal dimension of $h_0 = 0$, so the eigenvalue will be $e^{-2\pi i c/24}$. 
Under the modular $\mc S$ transformation, the characters transform into linear combinations of themselves. 
Therefore, if there is only one character, the character must remain invariant under the modular $\mc S$ transformation.
Combining both of these facts with the identity $(\mc S \mc T)^3 = \mathds 1$, it is easy to see that the central charge must be a multiple of 8 \cite{Schellekens:1992db},
\begin{equation}
  \label{eq:define8k}
  c = 8 k, \qquad k \in \mathbb Z^+.
\end{equation}
For a more in-depth approach using MLDEs, see \cite{Das:2023ybs}.

For $k = 1$, there is only one meromorphic CFT, the E$_{8,1}$ Wess-Zumino-Witten (WZW) model \cite{Witten:1983ar, Knizhnik:1984nr}.
The partition function of this CFT is
\begin{equation}
  Z(\tau,\bar \tau) = \left|j(\tau)^{\frac13}\right|^2,
\end{equation}
where $j(\tau)$ is the Klein-$j$ invariant.
The case $k=2$ contains two CFTs, $\rm SO(32)_1$ and $\rm E_{8,1} \otimes \rm E_{8,1}$, which are the two worldsheet CFTs for the heterotic string \cite{Gross:1984dd}.
The partition function for both CFTs is given by $Z(\tau,\bar \tau) = \left|j(\tau)^{\frac23}\right|^2$.
One should note that in addition, one needs the ghost fields on the worldsheet to cancel the central charge and therefore the conformal anomaly.
Schellekens finds a set of 71 (candidate) CFTs at $k=3$, or $c=24$ \cite{Schellekens:1992db}.
These CFTs have unit eigenvalue under the modular $\mc T$ transformation, which allows these CFTs to be chiral, which some authors take as another condition for a CFT to be meromorphic. 
In this paper, we will consider all single-character CFTs as meromorphic as a convention of nomenclature.
In general, the partition function of a meromorphic CFT is \cite{Mathur:1988na}
\begin{equation}
  Z(\tau,\bar\tau) = |\chi(\tau)|^2, \; \chi(\tau)  = j^{w_\rho}(j-1728)^{w_i} P_{w_\tau}(j),
\end{equation}
where $w_\rho= \{0,\frac13,\frac23\}$, $w_i = \{0,\frac12\}$, $w_\tau \in \mathbb Z$, and $P_{w_\tau}$ is a polynomial of degree $w_\tau$.

The partition function for $c=24$ meromorphic CFTs is given by $Z(\tau) = j(\tau) - 744 + \mc N$, where $\mc N$ controls the number of spin-one currents. 
If $\mc N = 0$, then the number of spin-one currents vanishes, and the corresponding CFT is the so-called Monster module.
This was the first realization of moonshine, and it initiated a fruitful exploration of the connections between number theory, group theory, and physics \cite{Frenkel:1989vertex, Conway:1979qga, Borcherds:1983sq, Dixon:1988qd, Bantay:1990tr, Borcherds:1992jjg, Chaudhuri:1995ee, Gaberdiel:2010ca, Gannon:2010yqa, Cheng:2011ay, Cheng:2012tq, Cheng:2014owa}.

Meromorphic CFTs play an important role in the construction and classification of RCFTs.
Using the novel coset construction, one can construct many CFTs, some of which do not correspond to any sets of minimal CFTs, \cite{Gaberdiel:2016zke, Das:2022slz, Das:2022uoe}.
The novel coset relates the characters of the coset CFTs to the character of the meromorphic CFT via a bilinear relation. 
Similarly, the conformal blocks of correlators in the coset CFTs can be reconstructed from the correlators of the secondaries in the meromorphic CFT \cite{Mukhi:2020sxt}.

Chiral meromorphic CFTs have also received interest due to their possible connection to exact holographic duals to AdS$_3$ gravity \cite{Witten:2007kt}.
Since the one-loop gravity partition function calculated from the heat kernel \cite{Giombi:2008vd}, Selberg methods \cite{Keeler:2018lza}, and Wilson spools \cite{Castro:2023bvo} fails to produce a modular invariant partition function, it may be possible to relate the corrections to the one-loop partition function to chiral meromorphic theories that are already modular invariant.
To obtain Einstein gravity from holography, we need to consider a large central charge, or large $k$.
For small $k$, the dual theory would be a highly stringy, quantum theory, which is not general relativity in a semiclassical limit.

Earlier attempts to calculate the R\'enyi entropy for some meromorphic CFTs on the torus have been perturbative, by using the small-interval expansion technique \cite{Das:2017vtp}.
This method uses the OPE of the twist operators to perturbatively compute the R\'enyi entropy for small intervals.
This allows one to compute the vacuum conformal block to a finite order by considering the one-point functions of vacuum secondaries.
Our method circumvents these shortcomings by nonperturbatively calculating all the conformal blocks of the twist two-point function using the Wro\'nskian method.

To calculate R\'enyi entropies for meromorphic CFTs, we first calculate the $\mathbb Z_N$ cyclic orbifold partition function.
The orbifold will introduce new primaries; therefore, the replicated CFT is no longer meromorphic.
The characters corresponding to the new primaries will turn out to be useful to normalize the conformal blocks for the twist two-point function, since on computing the fusion rules using the Verlinde formula, we find that the twist-antitwist fusion consists only of the vacuum. 

We demonstrate how to find the orbifold partition function for any $N$, but we will then focus on $N=2$ as an illustrative example.
We find that the $\mathbb Z_2$ cyclic orbifold of a meromorphic CFT yields a four-character CFT. 
The modular $\mc S$ matrix is calculated, and the fusion rules are derived for all $\mathbb Z_2$ cyclic orbifolds of a meromorphic CFT.  
We identify that these CFTs correspond to realizations of the toric code modular tensor category.
Furthermore, we find that for $k=1$, the conformal dimensions of two of the four primaries coincide and then go on to prove that their characters are equal.

Following this, we focus on the case $k=1$ with three characters.
This implies that the twist two-point function also has three conformal blocks. 
We then construct a third-order differential equation using the Wro\'nskian method, and constrain the coefficients of the differential equation using modular invariance and ellipticity of the coefficients and solutions.
The solutions of the differential equation are worked out in terms of Jacobi theta functions. 

Finally, we normalize the solutions by requiring that in the coincident limit the vacuum conformal blocks must reduce to the characters of the orbifold CFT. 
There is an ambiguity in the choice of normalization, which can be fixed by demanding the appropriate decompactification limit.
We find that the correlator is two-periodic along both cycles of the torus, which confirms and extends a result of \cite{Mukhi:2017rex}, where it is shown that the $N$th R\'enyi entropy of free fermions on a torus is $N$-periodic or periodic on the genus $N$ manifold, which is the $N$-fold cover of the torus.
In other words, the twist two-point correlator contains the information about the replica in its periodicities and is not just a consequence of the spin structures of fermions on a torus.
This result highlights the importance of attempting such calculations nonperturbatively, since this periodicity property is not obvious from a short-interval expansion approach.
We then compute the second R\'enyi entropy for a single interval on the torus for the $\rm E_{8,1}$ WZW model by taking the logarithm of the twist two-point function and comment on the $q$-expansion. 

The paper is organized as follows. 
In Sec. \ref{sec:review}, we review the Wro\'nskian method.
Then in Sec. \ref{sec:orbifiold}, we compute the characters, the fusion rules, and the number of conformal blocks in the replica meromorphic CFT.
Following that in Sec. \ref{sec:calc-renyi-entr} we use the Wro\'nskian method to develop a procedure to construct the differential equation satisfied by the conformal blocks of the two-point correlator of twist operators, finally calculating the R\'enyi entropy. 
Finally, in Sec. \ref{sec:discussion}, we make some concluding remarks and provide future directions.
In the appendixes, we list our definitions, conventions, and useful identities of elliptic functions and modular forms. 

\section{Review of computing torus correlators using Wro\'nskians}
\label{sec:review}

Correlation functions in conformal field theories can be expressed as a sum over holomorphic conformal blocks, since the Hilbert space of a conformal field theory is arranged into separate Verma modules, labeled by their respective primary field, which is the highest weight state of the symmetry group of the CFT.
The allowed conformal blocks correspond to the primaries generated by the fusion of the fields in the correlator and are therefore channel dependent.
The holomorphicity of the blocks refers to both the holomorphicity in the locations of the fields and the moduli of the Riemann surface.

Rational conformal field theories are CFTs which have a finite number of primary fields, so the correlators in such theories can be expressed as a finite sum over conformal blocks.
The correlator must be independent of the channel used to compute it; therefore, the conformal blocks must be transformed into each other when changing the channel.
This is known as crossing symmetry and is heavily exploited in the conformal bootstrap program of the classification of CFTs.

Let us begin by briefly reviewing \cite{Mathur:1988rx}, which introduces the procedure we will use to compute conformal blocks of correlators in RCFTs, which does not require the knowledge of intricate details of the CFT such as the specific null vectors.
Let us consider two-point functions of a primary field $\Phi(z,\bar z)$ and its conjugate $\overline\Phi(z,\bar z)$ on a torus with modular parameter $\tau$, 
\begin{equation}
  \label{eq:ConfBlock}
  \ev{\Phi(z_1,\bar z_1) \overline \Phi(z_2, \bar z_2)}_{\tau,\bar\tau} = \frac1{Z(\tau,\bar\tau)}\sum\limits_{i} d_i \mc F_i(z|\tau) \overline{{\mc F}_i}(\bar z|\bar \tau),
\end{equation}
where $d_i$ counts the number of primaries with the same conformal weight $h_i$, i.e., the degeneracy of the primary.
One can readily adapt our methods to non-diagonal CFTs where $d_i$ is promoted to a matrix, but we will consider diagonal CFTs.

Some properties of the conformal blocks $\mc F_i$ that will be useful in constructing a differential equation for them are listed below.
\begin{enumerate}
\item The conformal blocks transform into each other under the periodicity conditions:
  \begin{equation}\label{eq:monmat}
    \begin{split}
      \mc F_i(z+1|\tau) &= \sum_jM_{ij}^{(1)}\mc F_j(z|\tau), \\
      \mc F_i(z+\tau|\tau) &= \sum_j M_{ij}^{(\tau)}\mc F_j(z|\tau),
    \end{split}
  \end{equation}
  where $z=z_1-z_2$ due to translation invariance and $M^{(1)}, M^{(\tau)}$ are constant matrices known as the monodromy matrices for each cycle, respectively.
  
\item It is possible to diagonalize $M^{(1)}$ by choosing an appropriate basis of conformal blocks.
  Then the eigenvalues turn out to be phases $e^{2\pi i (h_\beta - h_{\beta'})}$, with $h_\beta, h_{\beta'}$ being the conformal dimensions of the intermediate primaries in the channel:
  
  \begin{figure}[!htbp]
    \centering
    \begin{tikzpicture}[scale=0.85]
      \draw (0,0) circle(1cm);
      \draw (-2,2) node [anchor=south]{$\Phi$} -- (140:1) node [anchor=east] {$z_1$};
      \draw (2,2) node [anchor=south]{$\bar \Phi$} -- (40:1) node [anchor=west] {$z_2$};
      \node at (0,1.22) {$\Phi_\beta$};
      \node at (0.1,-1.3) {$\Phi_{\beta'}$};
    \end{tikzpicture}
  \end{figure}
  
  In other words, this is the $z_1 \to z_1+1$ eigenstate basis. 
  This requires the fusion rule $\Phi \otimes \Phi_\beta = \Phi_{\beta'}$ to be nonzero.
  This is called the ``projection channel.''
  
\item Modular invariance implies the eigenvalues of $M^{(1)}$ and $M^{(\tau)}$ are the same.
  
\item Going around $z=0$ can be achieved by considering the loop $z\to z+1\to (z+1)+\tau \to (z+1+\tau)-1 \to (z+1+\tau-1)-\tau$.
  Therefore, the linear transformation $M^{(1)}M^{(\tau)}(M^{(1)})^{-1}(M^{(\tau)})^{-1}$ acts on the conformal blocks when circling the coincident point.
  Choosing the basis where this matrix is diagonal corresponds to the channel:
  
  \begin{figure}[!htbp]
    \centering
    \begin{tikzpicture}[scale=0.85]
      \draw (0,-1) circle(1cm);
      \draw (-1,2) node [anchor=south]{$\Phi_1$} -- (0,1);
      \draw (1,2) node [anchor=south]{$\Phi_2$} -- (0,1);      
      \draw (0,1) -- (0,0) node [midway, anchor=west] {$\Phi_{\beta}$};
      \node at (0.2,-2.3) {$\Phi_{\beta'}$};
    \end{tikzpicture}
  \end{figure}
  
  In other words, this channel corresponds to the $(z_1-z_2)\to e^{2\pi i}(z_1-z_2)$ eigenstate basis.
  This is called the ``OPE channel.''
\item The conformal blocks must transform into each other under modular transformations,
  \begin{equation}
    \begin{split}
      \mc F_i(z|\tau + 1) &= \sum_j\mc T_{ij}\mc F_j(z|\tau), \\    
      \mc F_i\left(\tfrac z\tau\left|-\tfrac1\tau\right.\right) &= \sum_j \mc S_{ij}\mc F_j(z|\tau).
    \end{split}
  \end{equation}
\end{enumerate}

This suggests that the computation for the conformal blocks of the two-point function will involve an $N$th order differential equation in $z\equiv z_1-z_2$ whose $n$-independent solutions are the $n$ conformal blocks that define the correlator,
\begin{equation}
  \partial^n \mc F + \sum_{i=0}^{n-1}\phi_i(z,\tau) \partial^i \mc F = 0.
\end{equation}
This can be motivated by  defining Wro\'nskians with the $n$ linearly independent conformal blocks,
\begin{equation}
  W_k = \det
  \begin{pmatrix}
    \mc F_1 & \dots & \mc F_n \\
    \partial \mc F_1 & \dots &  \partial \mc F_n \\
    \vdots & \dots & \vdots \\
    \partial^{k-1} \mc F_1 & \dots &  \partial^{k-1} \mc F_n \\
    \partial^{k+1} \mc F_1 & \dots &  \partial^{k+1} \mc F_n \\
    \vdots & \dots & \vdots \\
    \partial^n \mc F_1 & \dots &  \partial^n \mc F_n \\
  \end{pmatrix}, 
\end{equation}
with the $k$th derivative removed from the matrix to make it a square matrix.
We can derive the properties of $W_k$ from the properties of the conformal blocks $\mc F$:
\begin{enumerate}
\item Under a change in the basis of $\mc F$'s, the Wro\'nskian is multiplied by a $z$-independent constant.

\item The Wro\'nskian is invariant under $M^{(1)}$ and $M^{(\tau)}$.

\item The Wro\'nskian is a single-valued meromorphic function on the torus with poles only at $z=0$; i.e., the Wro\'nskian must be an elliptic function. 

\item The Wro\'nskian transforms under $\mc T$ and $\mc S$ as follows:
  \begin{equation}
    \label{eq:modWronski}
    \begin{aligned}
        W_k &\to (\det \mc T) W_k,\ \text{and}\\
        W_k &\to \tau^{\frac{n(n+1)}{2}-k}(\det \mc S) W_k.    
    \end{aligned}
  \end{equation}

\end{enumerate}

A useful property of the Wro\'nskian is $W_{n-1}= W_n'$, whose proof is straightforward in terms of wedge products.
Let us define $u_k = \partial^k \mc F_i \dd x^i$.
Then
\begin{equation}
  W_k = u_0 \wedge u_1 \wedge \dots \wedge u_{k-1} \wedge u_{k+1} \wedge \dots \wedge u_n. 
\end{equation}
Taking the derivative of $W_n$, using the fact that $\partial u_i = u_{i+1}$, we have
\begin{equation}
  \begin{split}
    \partial W_n
    &= u_1 \wedge u_1 \wedge \dots \wedge u_{n-1} + u_0 \wedge u_2 \wedge u_2\wedge \dots \wedge u_{n-1}\\
    &\quad + \dots + u_0 \wedge u_1 \wedge \dots \wedge u_{n-2} \wedge u_n\\
    &= u_0 \wedge u_1 \wedge \dots \wedge u_{n-2} \wedge u_n = W_{n-1}, 
  \end{split}
\end{equation}
by the antisymmetry of the wedge product, which completes the proof. 
We will use this construction to create the differential equation for conformal blocks. 
Since the wedge product distributes over addition, we have
\begin{equation}
  \begin{split}
    &W_{n-1} - \frac{W_{n-1}}{W_n} W_n\\
    &= u_0\wedge\dots\wedge u_{n-2}\wedge\left(u_n- \frac{W_{n-1}}{W_n} u_{n-1}\right) = 0.
  \end{split}
\end{equation}
For this to be zero, the terms in the wedge product must be linearly dependent, so we express the last term as a linear combination of $u_0,\dots, u_{n-2}$, giving us
\begin{equation}
  \begin{split}
    &\partial^n \mc F_i -\frac{W_{n-1}}{W_n}\partial^{n-1} \mc F_i+\sum_{r=0}^{n-2} \phi_r(z,\tau) \partial^r \mc F_i\\
    &= \left(\partial^n  + \sum_{r=0}^{n-1} \phi_r(z,\tau) \partial^r\right) \mc F_i = 0,
  \end{split}
\end{equation}
with 
\begin{equation}  
  \phi_{n-1} = -\frac{W_{n-1}}{W_n} = - \frac{W_n'}{W_n}.
\end{equation}
Similarly, we can construct each $\phi_k$ using similar arguments:
\begin{equation}
  \begin{split}
    W_k &= u_0 \wedge \dots \wedge u_{k-1}\wedge u_{k+1} \wedge \dots \wedge u_n\\
    & = u_0 \wedge \dots \wedge u_{k-1}\wedge u_{k+1} \wedge \dots \wedge \left(\sum_r^{n-1} \phi_r u_r\right)\\ 
    &= \sum_r^{n-1}\phi_r u_0 \wedge \dots \wedge u_{k-1}\wedge u_{k+1}\wedge \dots \wedge u_{n-1}\wedge u_r \\
    &= \phi_k  u_0 \wedge \dots \wedge u_{k-1}\wedge u_{k+1}\wedge \dots \wedge u_{n-1}\wedge u_k\\ 
    &= (-1)^{n-k}\phi_k(z,\tau) W_n,\\
    \implies& \phi_k(z,\tau) = (-1)^{n-k} \frac{W_k}{W_n}.
  \end{split}
\end{equation}
Since $W_k$'s are meromorphic functions with a pole at $z=0$, $\phi_r$'s are also meromorphic with poles at $z=0$ and at the zeros of $W_n$. 
Since $W_n$ is elliptic, it must have an equal number of poles and zeros.
\footnote{Let $f(z)$ be an elliptic function, i.e., meromorphic and periodic under $z\to z+1$ and $z \to z+\tau$.
To count the number of poles and zeros, one can use the Cauchy argument principle: $\frac1{2\pi i}\oint_{\mc C} \frac{f'(z)}{f(z)}dz = n_{\rm zeros} - n_{\rm poles}$. 
Since $f$ is doubly periodic, so is $f'/f$.
Choosing a contour $\mc C$ that encloses all the zeros and poles on the torus, i.e, along the parallelogram of $\mathbb C/(\mathbb Z + \tau \mathbb Z)$: $z\to z+1 \to z+1+\tau \to z+\tau \to z$, the integral must vanish by periodicity, and therefore, the number of zeros must be equal to the number of poles.}
The behavior of $\phi_r$ near $z=0$ or any other pole is a maximum singularity of $z^{r-n}$, which can be derived from a power series ansatz of the solution. 
We can also derive the modular properties of $\phi_r$, being
\begin{equation}
  \begin{split}
    \phi_r\left(\tfrac{z}{c\tau+d}, \tfrac{a\tau + b}{c\tau + d}\right)&= (c \tau + d)^{n-r}\phi_r(z,\tau),\\
    \phi_r(-z,\tau) &= (-1)^{n-r}\phi_r(z,\tau).
  \end{split}
\end{equation}
One can now use the behavior of $\phi_r$'s and $W_r$'s at the zeros and poles that are in terms of the conformal dimensions of the fields in the correlator and fields in the fusion rules to determine the differential equation.
For correlators with a low number of conformal blocks, this is tractable and is exactly solvable.

\section{Orbifolds of meromorphic CFTs}
\label{sec:orbifiold}

To calculate the $N$th R\'enyi entropy using the replica trick, the CFT must be replicated $N$ times.
These replicas are connected by branch cuts that represent the regions of interest, with the twist operators located at the end points of the intervals.
Taking the size of all the intervals to zero, we recover a tensor product CFT, with a cyclic $\mathbb Z_N$ symmetry, which we must remove with a $\mathbb Z_N$ cyclic orbifold.
This will modify the theory by introducing twisted sectors and new primaries associated with them and, therefore, new sets of fusion rules.

We also need to be able to construct the new characters in the orbifold CFT.
The coincident limit of the vacuum conformal blocks of a two-point correlator on the torus yield characters, 
\begin{equation}
  \begin{split}
    \ev{\phi(z,\bar z)\overline\phi(0,0)}_{\tau,\bar\tau} &= \frac1{Z(\tau,\bar\tau)}\sum_i d_i |\mc F_i(z|\tau)|^2,\\
    \chi_i(\tau) &= \lim_{z\to 0} z^{2h_\phi}\mc F_i(z|\tau) =
    \begin{tikzpicture}[baseline=(current bounding box.center)]
      \draw  (0,1) node[anchor=south] {$\mathds 1$}  -- (0,0.5);
      \draw (0,0) circle (0.5);
      \node at (0,-0.7) {$i$};
    \end{tikzpicture}\;,
  \end{split}
\end{equation}
where $i$ labels a primary which runs around the torus loop.
This allows us to normalize the conformal blocks, since the differential equation cannot tell us about the normalization.

The orbifold partition is modular invariant, and therefore, the twisted sectors must transform into each other under modular transformations to keep the sum modular invariant.
Consider a discrete group $G$, and group elements $g, h \in G$.

The orbifold partition function is constructed as
\begin{equation}
  Z_{\text{orb}}(\tau,\bar \tau) = \frac1{|G|}\sum_{g,h \in G} Z_{g,h}(\tau,\bar\tau). 
\end{equation}
If $G$ is non-Abelian, $g$ and $h$ must commute under the group action.
The twisted sectors transform under the action of the modular group as
\begin{equation} \label{eq:modtwist}
    \begin{aligned}
        \mc T Z_{g,h}(\tau,\bar\tau) &= Z_{g,gh}(\tau,\bar\tau),\ \text{and} \\
        \mc S Z_{g,h}(\tau,\bar\tau) &= Z_{h,g}(\tau,\bar\tau). 
    \end{aligned}
\end{equation}
To construct an orbifold, one can project to a $G$-invariant subspace,
\begin{equation}
  Z_{\text{proj}} (\tau,\bar\tau) = \frac1{|G|} \sum_{h\in G} Z_{1,h} (\tau,\bar\tau), 
\end{equation}
such that the partition function is still periodic in one cycle, and then use modular invariance to sum over all $g$ that commutes with $h$. 
A useful result to compute the $\mathbb Z_N$ cyclic orbifold partition functions for $N$ prime is
\begin{equation}\label{eq:ZNmodinvConst}
  Z_{\text{orb}}(\tau,\bar\tau) = \left(1+\sum_{m=0}^{N-1}\mc T^m \mc S\right)Z_{\text{proj}}(\tau,\bar\tau) - Z_{0,0}(\tau,\bar\tau). 
\end{equation}
Here, the group element $0$ represents the identity element of $\mathbb Z_N$ and $N$ has to be a prime number, since $\mc T^m \mc S$ produces every twist exactly once as a consequence of $\mathbb Z_N$ being simple (the only non-trivial subgroup is itself) for prime $N$, and the untwisted sector is subtracted to avoid overcounting.

Using this construction, we can write an expression for any $\mathbb Z_N$ orbifold CFT, by first constructing $Z_{\text {proj}}$ using some physical arguments \cite{Klemm:1990df}.
First, consider the CFT copied $N$ times.
This will correspond to the untwisted sector, whose partition function will be simply $Z(\tau,\bar \tau)^N$.
We also have to account for the tensor product states $\Phi=\prod\limits_{i=1}^N \phi_i$.
There are $N-1$ states equivalent up to cyclic permutations which are all identical when $\phi_i$ correspond to identical states in each copy, which have not been counted yet, so we have to add them.
The corresponding characters for these tensor product primaries can be written as follows:
\begin{equation}
  \chi_\Phi (\tau) =\Tr(q^{N L_0 - N \frac{c}{24}}) = \chi_\Phi(N \tau). 
\end{equation}
Finally, we have to divide by the order of the group $\left|\mathbb Z_N\right| = N$, since there is only one linear combination that corresponds to the totally symmetrized linear combination of states.
So we have
\begin{equation} \label{eq:ZNproj}
  Z_{\text{proj}}(\tau,\bar\tau) = \frac1N\left(Z(\tau,\bar \tau)^N  + (N-1) Z(N \tau, N \bar \tau)\right).
\end{equation}

Now we can use \eqref{eq:ZNmodinvConst} to write down the expression for the orbifold partition function.
First, we compute the action of the modular group generators, 
\begin{equation}\label{eq:ZNtwistSect}
  \begin{aligned}
    \mc T^m \mc S \, Z(N \tau, N \bar \tau) &= Z\left(-\tfrac{N}{\tau+m},-\tfrac{N}{\bar \tau+m}\right) \\
    &= Z\left(\tfrac{\tau + m}{N},\tfrac{\bar \tau + m}{N}\right).
  \end{aligned}
\end{equation}
So, we can  express the $\mathbb Z_N$ orbifold partition function:
\begin{equation}\label{eq:ZNorbPF}
  \begin{split}
    Z_{\text{orb}}(\tau,\bar\tau)
    ={}& \frac1N Z(\tau,\bar \tau)^N 
    + \frac{N-1}{N}Z(N \tau, N\bar \tau)\\
    {}&+ \frac{N-1}{N}\sum_{m=0}^{N-1}Z\left(\tfrac{\tau + m}{N},\tfrac{\bar \tau + m}{N}\right).
  \end{split}
\end{equation}
This result has been generalized to any natural number $N$ by \cite{Haehl:2014yla}
\begin{equation}
  \begin{split}
    &Z_{\rm orb} (\tau,\bar\tau)\\
    &= \frac1N \sum_{r,s=1}^N
    Z\left(\tfrac{\gcd(N,r)}{N}\left(\tfrac{\gcd(N,r)}{\gcd(N,r,s)} \tau + \kappa(r,s)\right) \right)^{\gcd(N,r,s)},\\
    &\kappa(r,s) = \min\left\{0,1,\dots, \tfrac{N}{\gcd(N,r)}-1\right\}\\
    &\text{such that} \left(\kappa(r,s) r  - \tfrac{\gcd(N,r)s}{\gcd(N,r,s)}\right) = 0 \mod N. 
  \end{split}
\end{equation}
This can also be written in terms of ``square-free'' Hecke operators and the Euler's totient function $\varphi$ \cite{Takayanagi:2022xpv},
\begin{equation}
  Z_{\rm orb} = \sum_{d|N} \frac{\varphi(N/d)}{d} T_{N/d}^{sf}\left(Z(\tau)^d\right),
\end{equation}
where the square-free Hecke operators are recursively defined in terms of Hecke operators $T_k$,
\begin{equation}
  \begin{split}
    T_kZ(\tau) &= \frac1k\sum_{i|k}\sum_{j=0}^{i-1}Z\left(\frac{k\tau}{i^2}+\frac ji\right),\\
    T_{k}^{sf}Z(\tau) &= T_kZ(\tau) - \sum_{a>1,\, a^2|k}\frac1{a^2} T_{k/a^2}^{sf} Z(\tau).
  \end{split}
\end{equation}
For prime $k$, the square-free and the regular Hecke operators coincide.

\subsection{Calculating the twist operator fusion rules and number of conformal blocks}
\label{sec:calc-twist-oper}

In this section, we will calculate the fusion rules for a $\mathbb Z_2$ cyclic orbifold of a meromorphic CFT.
The reason we restrict ourselves to $N=2$ is because the number of characters in the orbifold CFT grows with $N$, which means that the resulting number of conformal blocks for the correlator of interest grows with $N$ as well, and therefore the corresponding order of the differential equation. 
In principle, this can be worked out for any $N$, but for illustrative and calculation purposes, we shall stick to $N=2$.

The $\mathbb Z_2$-invariant projection of the partition function is 
\begin{equation}
  \begin{split}
    Z_{\text{proj}}(\tau,\bar\tau)
    &= \frac12 (Z_{0,0}(\tau,\bar \tau) + Z_{0,1}(\tau,\bar \tau))\\
    &= \frac12(Z(\tau,\bar \tau)^2 + Z(2\tau,2\tau)),
  \end{split}
\end{equation}
where, using \eqref{eq:modtwist}, one can identify $Z_{0,0}(\tau,\bar \tau) = Z(\tau,\bar\tau)^2$ and $Z_{0,1}(\tau,\bar \tau) = Z(2\tau,2\bar\tau)$ since $Z(\tau,\bar\tau)^2$ must be invariant under $\mc S$.
It is clear to see that $Z_{\text{proj}}$ is $\mc T$ invariant.
Similarly $Z_{0,0}(\tau,\bar \tau)-Z_{0,1}(\tau,\bar \tau)$ is also $\mc T$ invariant, which will come in handy later.

The full orbifold partition function is simply
\begin{equation} \label{eq:ZorbDef}
  \begin{split}
    &Z_{\text{orb}}(\tau,\bar\tau)\\
    &= \frac12 (Z_{0,0}(\tau,\bar \tau)+ Z_{0,1}(\tau,\bar \tau) + Z_{1,0}(\tau,\bar \tau) + Z_{1,1}(\tau,\bar \tau)) \\
    &= \frac12 \left(Z(\tau,\bar \tau)^2+ Z(2\tau,2\bar \tau) + Z\left(\tfrac\tau2,\tfrac{\bar\tau}2\right) + Z\left(\tfrac{\tau+1}2,\tfrac{\bar\tau+1}2\right)\right),
  \end{split}
\end{equation}
where we have used
\begin{equation}
  \begin{split}
    Z_{1,0}(\tau,\bar\tau)
    &= \mc S Z_{0,1}(\tau,\bar\tau) = Z\left(-\tfrac2\tau,-\tfrac2{\bar\tau}\right) = Z\left(\tfrac\tau2,\tfrac{\bar\tau}2\right),\\
    Z_{1,1}(\tau,\bar\tau)
    &= \mc T Z_{1,0}(\tau,\bar\tau) =  Z\left(\tfrac{\tau+1}2,\tfrac{\bar\tau+1}2\right).
  \end{split}
\end{equation}

Now, since the parent CFT is a meromorphic CFT, we can write the parent partition function as the absolute value squared of the vacuum character alone,
\begin{equation} 
  Z(\tau,\bar \tau) = |\chi(\tau)|^2,
\end{equation}
such that $\mc S \chi(\tau) = \chi(\tau)$  and $\mc T \chi(\tau) = e^{-2\pi i c/24}\chi(\tau)$.
We can therefore write the characters of each twisted sector as
\begin{equation}
  Z_{i,j}(\tau,\bar\tau) = \left|\chi_{i,j}(\tau)\right|^2.
\end{equation}
However, characters must be in a $\mc T$ eigenbasis to be physical, and they also must have only one vacuum character.
Currently, in the current twisted basis, the two modular transformations are
\begin{equation}
  \begin{split}
    \mc T_{\text{twist}} \begin{pmatrix} \chi_{0,0} \\ \chi_{0,1} \\\chi_{1,0} \\\chi_{1,1} \\\end{pmatrix} &=
    \begin{pmatrix} \lambda_0 & 0 & 0 & 0 \\ 0 & \lambda_0 & 0 & 0\\0 & 0 & 0 & \lambda_1\\0 & 0 & \lambda_1 & 0 \end{pmatrix}
    \begin{pmatrix} \chi_{0,0} \\ \chi_{0,1} \\\chi_{1,0} \\\chi_{1,1} \\\end{pmatrix},\\
    \mc S_{\text{twist}} \begin{pmatrix} \chi_{0,0} \\ \chi_{0,1} \\\chi_{1,0} \\\chi_{1,1} \\\end{pmatrix} &=
    \begin{pmatrix} 1 & 0 & 0 & 0 \\ 0 & 0 & 1 & 0\\0 & 1 & 0 & 0\\0 & 0 & 0 & 1 \end{pmatrix}
    \begin{pmatrix} \chi_{0,0} \\ \chi_{0,1} \\\chi_{1,0} \\\chi_{1,1} \\\end{pmatrix},\\
  \end{split}
\end{equation}
which follows from \eqref{eq:modtwist}.
Clearly, this basis is not a $\mc T$ eigenbasis.
Also, this basis suffers from the fact that $\mc S_{(0,0),i} = 0$ for $i\neq(0,0)$, which yields indeterminate fusion rules.
This is a consequence of the fact that both $\chi_{0,0}(\tau)$ and $\chi_{0,1}(\tau)$ have the same leading behavior at $\tau = i\infty$, or the same $\mc T$ eigenvalue, which makes one to have the incorrect conclusion that both correspond to the vacuum character in the orbifold theory, which is clearly incorrect.

Luckily, both of these problems can easily be rectified by making the change of basis:
\begin{equation}\label{eq:Tbasis}
  \begin{split}
    \begin{pmatrix} \chi_0 \\ \chi_1 \\ \chi_2 \\ \chi_3 \end{pmatrix}
    &\equiv A \begin{pmatrix} \chi_{0,0} \\ \chi_{0,1} \\\chi_{1,0} \\\chi_{1,1} \\\end{pmatrix}
    =  \frac12 \begin{pmatrix} \chi_{0,0}+\chi_{0,1} \\ \chi_{0,0}-\chi_{0,1} \\\chi_{1,0}+\chi_{1,1} \\\chi_{1,0}-\chi_{1,1} \\\end{pmatrix}\\
    \implies A &= \frac12
    \begin{pmatrix}
      1 & 1 & 0 & 0 \\
      1 & -1 & 0 & 0 \\
      0 & 0 & 1 & 1 \\
      0 & 0 & 1 & -1 \\
    \end{pmatrix},
  \end{split}
\end{equation}
which diagonalizes $\mc T = A \, \mc T_{\text{twist}} \, A^{-1}$ and yields
\begin{equation}  \label{eq:smat4}
  \mc S = A \,\mc S_{\text{twist}}\, A^{-1}=
  \frac12
  \begin{pmatrix}
    1 & 1 & 1& 1\\
    1 & 1 & -1& -1\\
    1 & -1 & 1& -1\\
    1 & -1 & -1& 1\\
  \end{pmatrix}.      
\end{equation}
The characters reproduce the result derived in \cite{Borisov:1997nc}, which computes the characters for $\mathbb Z_N$ cyclic orbifold CFTs. 

To determine the fusion rules, one can use Verlinde's formula \cite{Verlinde:1988sn}
\begin{equation}  
  \mc N_{ij}^{~~k} = \sum_m \frac{\mc S_{im}\mc S_{jm}\mc S_{mk}^{-1}}{\mc S_{0m}},  
\end{equation}
where $\mc N_{ij}^{~~k} =1$ implies that the OPE of the primaries $\phi_i$ and $\phi_j$ contains $\phi_k$ and its descendants.
The fusion rules for the $\mathbb Z_2$ cyclic orbifold of meromorphic CFTs are therefore,
\begin{equation}
  \begin{split}
    \mc N^{0} =
    \begin{pmatrix}
      1 & 0 & 0 & 0\\
      0 & 1 & 0 & 0\\
      0 & 0 & 1 & 0\\
      0 & 0 & 0 & 1\\
    \end{pmatrix},\quad 
    \mc N^{1} =
    \begin{pmatrix}
      0 & 1 & 0 & 0\\
      1 & 0 & 0 & 0\\
      0 & 0 & 0 & 1\\
      0 & 0 & 1 & 0\\
    \end{pmatrix},\\
    \mc N^{2} =
    \begin{pmatrix}
      0 & 0 & 1 & 0\\
      0 & 0 & 0 & 1\\
      1 & 0 & 0 & 0\\
      0 & 1 & 0 & 0\\
    \end{pmatrix},\quad
    \mc N^{3} =
    \begin{pmatrix}
      0 & 0 & 0 & 1\\
      0 & 0 & 1 & 0\\
      0 & 1 & 0 & 0\\
      1 & 0 & 0 & 0\\
    \end{pmatrix},
  \end{split}  
\end{equation}
where $0,\dots, 3$ correspond to the characters defined in the order of \eqref{eq:Tbasis}.
This turns out to be the fusion class $\mc A^{(1)}_3$ as in \cite{Christe:1988xy}
\begin{equation}  
  \mc A^{(1)}_3:\quad \mc N_{011} = \mc N_{022} = \mc N_{033} = \mc N_{123} = 1.
\end{equation}

We can now compute the number of conformal blocks.
We see that a primary fusing with itself only yields the vacuum, and the number of conformal blocks is given by
\begin{equation}  
  \mc N = \sum_j \mc N_{0j}^{~~j} = 4.
\end{equation}
Therefore, we have four conformal blocks in the two-point correlator in a $\mathbb Z_2$ orbifold meromorphic CFT.
This implies that the conformal blocks are solutions of a fourth-order differential equation.

Let us note the conformal dimensions of the four primary fields corresponding to each of the characters, from the leading exponent of each character:
\begin{equation}
  \begin{aligned}
    \chi_0(\tau) \sim{}& q^{-\frac{2k}3} \equiv q^{-\frac{c}{24}}, 
    &\chi_1(\tau) \sim{}& q^{-\frac{2k}3+1} \equiv q^{h_1-\frac{c}{24}}, \\
    \chi_2(\tau) \sim{}& q^{-\frac k6} \equiv q^{h_2-\frac{c}{24}},
    &\chi_3(\tau) \sim{}& q^{-\frac k6 +\frac12} \equiv q^{h_3-\frac{c}{24}},
  \end{aligned}
\end{equation}
we find that $c = 16 k,$ $h_1=1,$ $h_2=\frac k2,$ and $h_3=\frac k2 + \frac12$.
This suggests that $\chi_0$ is the vacuum character, and $\chi_2$ is the character corresponding to the twist operator.
The primaries $\chi_1$ and $\chi_3$ correspond to the spin-one current and the twist operator corresponds to the current, respectively \cite{Borisov:1997nc}.
Note that the twist operator is self-conjugate.
This is consistent since the twist operator takes one to the next Riemann sheet, and the antitwist operator takes one to the previous Riemann sheet.
Since we are working with a $\mathbb Z_2$ cyclic orbifold, there are only two Riemann sheets, and therefore moving to the next or previous sheet is the same operation. 

For completeness, let us compute the Wro\'nskian index for these orbifold CFTs.
For $k\geq2$, we have
\begin{equation}
  \begin{aligned}
    \ell &= 6\left[ \frac{n(n-1)}{12} -\sum_{i=0}^{n-1}\left(h_i - \frac{c}{24}\right) \right] \\
    &= 6\left(\frac{5k}3 - \frac12\right),
  \end{aligned}
\end{equation}
where $n=4$ is the number of characters.
For instance, when $k=2$, it follows that $\ell = 17$, and for $k=3$, $\ell = 27$, and so forth.
This suggests that considering cyclic orbifolds of simple low-character CFTs is a useful way of generating CFTs with large Wro\'nskian indices and characters.

This class of CFTs falls under the so-called ``toric code modular tensor category'' \cite{Rowell:2007dge}, since they have a vanishing topological central charge ($c$ mod 8) and are of rank 4; i.e., they have four primaries and satisfy the appropriate modular properties.

\subsection{Degeneracy at \texorpdfstring{$k=1$}{}}
\label{sec:degenaracy-at-k=1}

Note that when $k=1$, both $\chi_1$ and $\chi_3$ correspond to characters of primaries of conformal dimension $h_1=h_3 =1$.
Our goal now is to demonstrate that this represents a physical scenario, specifically that the two characters are identical.

The parent CFT is the E$_{8,1}$ WZW model with $c=8$, and the partition function is given by
\begin{equation}
  Z(\tau,\bar \tau) = |j(\tau)^{\frac13}|^2,
\end{equation}
where $j(\tau)$ is the Klein-$j$ invariant [cf. \eqref{eq:jdef1}, \eqref{eq:jdef2}].
The $q$-expansion of $j(\tau)$ is given by, 
\begin{equation}\label{eq:jdef}
  \begin{split}
    j(\tau)
    &= \frac{(\vartheta_2(\tau)^8 + \vartheta_3(\tau)^8+\vartheta_4(\tau)^8)^3}{8\eta(\tau)^{24}}\\
    &= \frac1q + 744 + 196884q + \mc O(q^2).
  \end{split}
\end{equation}
It is clear in this representation of the Klein-$j$ invariant that it is as a cube, whose cube root also has an integral $q$-expansion,
\begin{equation}
  j(\tau)^{\frac13}= q^{-\frac13}(1+248q + 4124q^2 + \mc O(q^3)), 
\end{equation}
where the coefficient 248 indicates the number of spin-one currents, which should be equal to the number of generators of the Lie group $\rm E_8$.

\begin{widetext}
  The $\mathbb Z_2$ cyclic orbifold partition function \eqref{eq:ZorbDef} is
  \begin{equation} 
    Z_{\text{orb}}=\frac12 \left(\left|j(\tau)^{\frac23}\right|^2+ \left|j(2\tau)^{\frac13}\right|^2 + \left|j\left(\tfrac\tau2\right)^{\frac13}\right|^2 + \left|j\left(\tfrac{\tau+1}2\right)^{\frac13}\right|^2\right),
  \end{equation}
  with the characters
  \begin{equation}\label{eq:chainj}
    \begin{split}
      \chi_0(\tau) &= \frac12(j(\tau)^{\frac23}+j(2\tau)^{\frac13}) =
      q^{-\frac23}+248 q^{\frac13}+ 35000 q^{\frac43} + \mc O(q^{\frac73}),\\
      \chi_1(\tau) &= \frac12(j(\tau)^{\frac23}-j(2\tau)^{\frac13}) =
      248 q^{\frac13} + 34752 q^{\frac43} + 1057504  q^{\frac73}+\mc O(q^{\frac{10}{3}}),\\
      \chi_2(\tau) &= \frac12\left(j\left(\tfrac\tau2\right)^{\frac13}+(-1)^{\frac13}j\left(\tfrac{\tau+1}2\right)^{\frac13}\right)=
      q^{-\frac16} + 4124 q^{\frac56} + 213126  q^{\frac{11}6}+\mc O(q^{\frac{17}{6}}),\\
      \chi_3(\tau) &=\frac12\left(j\left(\tfrac\tau2\right)^{\frac13} -(-1)^{\frac13}j\left(\tfrac{\tau+1}2\right)^{\frac13}\right)= 248 q^{\frac13} + 34752 q^{\frac43} + 1057504  q^{\frac73}+\mc O(q^{\frac{10}{3}}).  
    \end{split}
  \end{equation}
\end{widetext}
The coefficient $\left(-1\right)^{\frac13}$ of ${j\left(\frac{\tau+1}2\right)}^{\frac13}$ can be worked out by comparing the $q$-expansions, by ensuring that the coefficients are non-negative integers, and also by using the expressions relating the twisted sectors and characters found in \cite{Borisov:1997nc}.

For this to be a physical theory, the characters $\chi_1$ and $\chi_3$ must be equal, since they have the same $\mc T$ eigenvalue.
One can check using a computer algebra system that the $q$-expansions match up to arbitrary order, but that is not enough to check that the characters are equal.
A formal proof can be found by expressing the characters in terms of Jacobi theta functions (details can be found in Appendix \ref{app:elliptic-functions}).

The characters \eqref{eq:chainj} expressed in terms of theta functions are
\begin{equation}
  \begin{split}
    \chi_0(\tau) &= \frac{16\vartheta_3(\tau)^{16}- 31\vartheta_2(\tau)^8\vartheta_3(\tau)^4\vartheta_4(\tau)^4+ 16\vartheta_4(\tau)^{16}}{32\eta(\tau)^{16}},\\
    \chi_1(\tau) &= \chi_3(\tau) \\ &=\frac{\vartheta_2(\tau)^8\left(16\vartheta_3(\tau)^8 - \vartheta_3(\tau)^4\vartheta_4(\tau)^4 + 16 \vartheta_4(\tau)^8\right)}{32\eta(\tau)^{16}},\\
    \chi_2(\tau) &= \frac{(\vartheta_3(\tau)^8-\vartheta_4(\tau)^8)}{32\eta(\tau)^{16}}\\
    &\times\left(16\vartheta_3(\tau)^8 - 31\vartheta_3(\tau)^4\vartheta_4(\tau)^4 + 16 \vartheta_4(\tau)^8\right).
  \end{split}
\end{equation}
Thus, we have found the normalizations of the conformal blocks $\mc F_i(z|\tau)$, by demanding that the limit $z\to 0$ recovers the characters $\chi_i(\tau)$. 

Let us note the modular $\mc S$ matrix for these characters,
\begin{equation}
  \mc S \begin{pmatrix} \chi_0\\ \chi_1 \\ \chi_2 \end{pmatrix}
  =\frac12
  \begin{pmatrix}
    1&2&1\\1&0&-1\\1&-2&1 
  \end{pmatrix}
  \begin{pmatrix} \chi_0\\ \chi_1 \\ \chi_2 \end{pmatrix},
\end{equation}
which can be worked out using either the properties of theta functions or the properties of the Klein-$j$ invariant.
As a check, one can work out that the modular $\mc S$ matrix satisfies the following properties:
\begin{equation}
  \mc S^2 = \mathds 1,
  \quad \mc S^{\mathsf T}\begin{pmatrix}1&0&0\\0&2&0\\0&0&1\end{pmatrix}\mc S
  =\begin{pmatrix}1&0&0\\0&2&0\\0&0&1\end{pmatrix},
\end{equation}
the second of which correctly implies that the partition function is
\begin{equation}\label{eq:Zorb}
  Z_{\rm orb}(\tau,\bar \tau) = |\chi_0(\tau)|^2 + 2|\chi_1(\tau)|^2 + |\chi_2(\tau)|^2.
\end{equation}
Note that the $\mc S$ matrix is nonunitary and therefore cannot be used to compute the fusion rules in this state.
To do so, we must use the $4\times4$ matrix computed in \eqref{eq:smat4} to compute the fusion rules, treating both primaries separately. 

The Wro\'nskian index for $k=1$ is
\begin{equation}
  \ell = 6\left[ \frac{n(n-1)}{12} -\sum_{i=0}^{n-1}\left(h_i - \frac{c}{24}\right) \right] = 6,
\end{equation}
where $n=3$ is the number of characters, as we have now shown.
This is an explicit realization of an $\ell = 6$ three-character CFT with central charge 16. 

Let us briefly compare this CFT ($\frac{\rm E_{8,1}\otimes E_{8,1}}{\mathbb Z_2}$) with another $c=16$, $\ell =6$ CFT, namely $\rm SO(16)_1 \otimes E_{8,1}$.
\footnote{We thank Sunil Mukhi for bringing this to our attention.}
The latter has the same CFT data as the former, that is, two primaries with $h=1$ and a third primary with $h=\frac12$. 
However, since $\ell\neq 0$, just matching these CFT data is not enough to say that these two are the same CFTs. 
We can check this by comparing the characters of the two CFTs.
The characters of $\rm SO(16)_1 \otimes E_{8,1}$ are readily computed to be
\begin{equation}
  \begin{aligned}
    \chi_0 (\tau) &= \frac12 \left(\frac{\vartheta_3(\tau)^8}{\eta(\tau)^8} + \frac{\vartheta_4(\tau)^8}{\eta(\tau)^8} \right)j(\tau)^{\frac13} \\
    &= q^{-\frac23} + 368 q^{\frac13} + \mc O(q^{\frac43}),\\
    \chi_1 (\tau) &= \frac12 \frac{\vartheta_2(\tau)^8}{\eta(\tau)^8} j(\tau)^{\frac13} \\
    &= 128 q^{\frac13}+ 33792 q^{\frac43} + \mc O(q^{\frac73}),\\
    \chi_2 (\tau) &= \frac12 \left(\frac{\vartheta_3(\tau)^8}{\eta(\tau)^8} - \frac{\vartheta_4(\tau)^8}{\eta(\tau)^8} \right)j(\tau)^{\frac13} \\
    &= 16 q^{-\frac16}+ 4544 q^{\frac56}+ \mc O(q^{\frac{11}6}),
  \end{aligned}
\end{equation}
which clearly show that the two CFTs are inequivalent, despite having identical modular properties. 
Their modular $\mc S$ matrix and fusion rules also match, as a result of this CFT satisfying the conditions for the toric code modular tensor category.
This is not a problem since CFTs with nonzero Wro\'nskian index have movable poles in the MLDE satisfied by their characters \cite{Das:2023qns}.

\section{Calculating the R\'enyi entropy }
\label{sec:calc-renyi-entr}

As stated in the Introduction, to compute the second R\'enyi entropy we must consider the correlator of twist operators on the torus, 
\begin{equation}
  \ev{\sigma(z,\bar z)\sigma(0,0)}_{\tau,\bar\tau} = \frac1{Z_{\mathrm{orb}}(\tau,\bar\tau)}\sum_i d_i|\mc F_i(z|\tau)|^2. 
\end{equation}
Note that, as described in Sec. \ref{sec:calc-twist-oper}, the twist operator is self-conjugate when considering a $\mathbb Z_2$ cyclic orbifold, and hence we do not put a bar over the second twist operator.

The number of conformal blocks for this correlator for the second R\'enyi entropy is 4, as calculated prior, except for the $c=8$ meromorphic theory, the E$_{8,1}$ WZW model, where we have 3 conformal blocks, one of the blocks having a multiplicity of 2.

In this section, we shall work out the R\'enyi entropy of the E$_{8,1}$ WZW model as an illustrative example, while exploring general meromorphic CFTs along the way.  

\subsection{Constructing the differential equation for the conformal blocks}
\label{sec:constructingblocks}

First, let us make some comments on the $k\geq 2$ cases, with four distinct conformal blocks.
Since there are four conformal blocks, we know that the blocks have to satisfy a fourth-order differential equation:
\begin{equation}
  \partial^4 \mc F + \sum_{m=0}^3 \phi_m\partial^m \mc F = 0,
\end{equation}
where
\begin{equation}
  \phi_m = (-1)^{4-m} \frac{W_m}{W_4}.
\end{equation}
To work out the Wro\'nskian $W_4$, let us recall that the twist operators have a conformal dimension of $h_\sigma = \frac k2$.
All blocks are vacuum blocks, so they will all have the leading singularity $z^{-2h_\sigma} = z^{-k}$.
One can construct a linear combination to eliminate the leading singularity in the other two blocks, so the leading singularity for the second block will be $z^{-k+2}$.
The $z^{-k+1}$ term should be zero since that term would correspond to the one-point function of a current secondary, which should be zero due to charge conservation.
The third and fourth blocks can be constructed similarly such that the leading singularities are all unique, and will be $z^{-k+3}$ and $z^{-k+4}$.
Working out the Wro\'nskian, we have the leading singularity 
\begin{equation}
  W_4 \sim z^{3-4k}. 
\end{equation}
Since $W_4$ must be an elliptic function with poles only at $z=0$, the Wro\'nskian can be expressed as a polynomial in the Weierstra{\ss} $\wp$ function and its derivatives,
\begin{equation}
  W_4(z,\tau) = \alpha^{(4)}_{0}(\tau)+\sum_{l=0}^{4k-5}\alpha^{(4)}_{l+2}(\tau) \partial_z^l \wp(z|\tau).
\end{equation}

Similarly, we can compute the rest of the Wro\'nskians using similar arguments, where we find
\begin{equation}
  W_m \sim z^{m-4k-1}.
\end{equation}
This allows us to write the differential equation for the conformal blocks as follows:
\begin{equation}
  \sum_{m=0}^4 \left( \alpha^{(m)}_{0}(\tau)  + \!\!\!\!\sum_{l=0}^{4k-m-1} \alpha^{(m)}_{2+l}(\tau) \partial^l \wp(z|\tau) \right)\partial^m \mc F(z|\tau) = 0
\end{equation}
The coefficients $\alpha^{(m)}_i(\tau)$ will be modular forms. 
To fully work them out, we have to work out the number of zeros of the Wro\'nskian $W_4$ as a function of $\tau$ in the fundamental domain.

We first calculate the behavior of $W_4$ in the limit $\tau\to i\infty$ for a finite $z$. 
To do so, we have to work in the projection basis, 
\begin{equation}
  \begin{tikzpicture}[baseline=(current bounding box.center),scale=0.5]
    \draw (0,0) circle(1cm);
    \draw (-2,2) node [anchor=south]{$\Phi$} -- (140:1) node [anchor=east] {$z_1$};
    \draw (2,2) node [anchor=south]{$\bar \Phi$} -- (40:1) node [anchor=west] {$z_2$};
    \node at (0,1.355) {$\Phi_\beta$};
    \node at (0.1,-1.4) {$\Phi_{\beta'}$}; 
  \end{tikzpicture}
  \sim q^{|h_\beta - h_{\beta'}| - \frac c{24}} \mc F(z_{12}).
\end{equation}
Using the conformal weights of the primary fields in the $\mathbb Z_2$ cyclic orbifold and their fusion rules, we find that the decompactification limits of the blocks are $q^{\frac k2 - \frac23 k}$, $q^{\frac k2 - \frac23 k}$, $q^{\frac{k-1}2-\frac23 k}$, and $ q^{\frac{k-1}2 -\frac23 k}$. 
This implies $W_4 \sim q^{-1-\frac23 k}$ in the $\tau\to i \infty$ limit. 

The number of zeros of the Wro\'nskian is given by $\frac b{12}-a$ \cite{Mathur:1988rx}, where $W\sim q^a$ and $\mc S\, W \sim \tau^b W$.
This can be derived from the valence formula for a modular form of weight $k$ \eqref{eq:valence}. 
We have from the behavior of the blocks that $a = -1-\frac23$.
To find $b$, we can use the following argument.
Using \eqref{eq:modWronski}, we see that $W_4 \to \tau^6 (\det M^{(\mc S)}) W_4$.
Since all the conformal blocks are vacuum blocks, i.e., they reduce to characters in the coincident limit, the $\mc S$ matrix for the blocks $M^{(\mc S)}$ must be proportional to the $\mc S$ matrix of the characters:
\begin{equation}  
  \begin{split}
    \chi_i(\tau) &= \lim_{z\to 0} z^{2h_\sigma} \mc F_i(z|\tau), \\
    \mc S_{ij}\chi_j(\tau) &= \chi_i\left(-\tfrac1\tau\right) 
    = \lim_{z\to 0} \frac{z^{2h_\sigma}}{\tau^{2h_\sigma}} \mc F_i\left(\tfrac z\tau\left.|-\tfrac1\tau\right.\right) \\
    &= \tau^{-2h_\sigma} M^{(\mc S)}_{ij}\lim_{z\to 0} z^{2h_\sigma} \mc F_j( z|\tau),
  \end{split}
\end{equation}
which implies
\begin{equation}
  \label{eq:4}
  M^{(\mc S)} = \tau^{2h_\sigma }\mc S,\quad  \implies \det M^{(\mc S)} = \pm \tau^{2n h_\sigma},
\end{equation}
since $\mc S^2= 1$, and $n$ is the number of distinct characters.
Clearly, $n=4$ for $k\geq2$, so $W_4 \to \tau^{6+4k} W_4$ under the modular $\mc S$ transformation.  

Therefore, the number of zeros of $W_4$ in the fundamental domain is $\frac{6+4k}{12} - \left(-1-\frac23 k\right) = k+\frac32$.
Clearly, this implies that $\alpha^{(4)}_0$ must be expressed as $E_6(\tau)^3 f_{k}(\tau)$, where $f_k(\tau)$ is a modular form of weight $k$, since the zero of order $\tfrac32$ must originate due to a factor of $E_6(\tau)^3$.
This means that $\alpha^{(4)}_0$ must be a modular form of weight $18+k$.
Similarly, we can identify that the weight of $\alpha^{(m)}_{l+2}$ is $20 + k - m - l$.
This differential equation can now be solved on a case-by-case basis, using methods such as the Frobenius method to obtain a power series solution.
In practice, however, one might need additional information to constrain the differential equations for larger $k$, since the number of zeros and constraints from the power series recurrence relations alone may not be enough to fully determine the coefficients.
This may include knowing the behavior of the conformal blocks in the decompactified limit, i.e., the conformal blocks of the four  point functions on the plane where two of the four operators are the operators in the original torus two-point function, and the other two operators are the same primary, corresponding to the primary in the loop channel.
We shall see that for $k=1$ this will not be a problem. 

The $k=1$ case is distinct, since we have three conformal blocks, whose leading singularities are $z^{-1}, z^{1}, z^{2}$, since the term $z^0$ is missing for the same reason as in the other cases.
The leading singularity of the Wro\'nskian $W_3$ is 
\begin{equation}
  W_3 \sim z^{-1}. 
\end{equation}
However, there are no elliptic functions with only one simple pole on the torus; therefore, $W_3$ must be a constant in $z$, which implies that the leading singularity for the third block is, in fact, $z^3$, instead of $z^2$.
The limit $\tau\to i\infty$ behavior for the blocks is $q^{-\frac23},q^{-\frac16},q^{-\frac16}$, which shows that the Wro\'nskian behaves as $W_3 \sim q^{-1}$. 

Let us now work out the differential equations up to constants in $z$, focusing on the $k=1$ case for simplicity.
Since $W_{n-1}= W_n'$, we have $W_2 = 0$.
The differential equation works out to be
\begin{equation}
  \begin{split}
    &\partial^3\mc F + (\alpha_1(\tau) \wp(z|\tau)+\alpha_2(\tau))\partial \mc F\\
    &\qquad + (\beta_1(\tau) \wp'(z|\tau)+\beta_2(\tau)) \mc F  = 0.
  \end{split}
\end{equation}

Since $n=3$, and $h_\sigma = \frac12$, we have $b=6$ and $a=-1$, and therefore, the number of zeros of $W_3(\tau)$ in the fundamental domain is $\frac{6}{12}+ 1 = \frac32$ [cf. \eqref{eq:valence}].
We find that we have a half integer number of zeros.
The zero of order half implies that the Wronskian contains a factor of the modular form $E_6(\tau)$, since $E_6(\tau)$ is the modular form with a zero of order half, which is located at $\tau = i$.
The location of the order-1 zero can be anywhere in the fundamental domain of $\mathrm{PSL}_2(\mathbb Z)\backslash \mathbb H$, but must be due to a modular form of weight 12, all of which can be written as $m E_4(\tau)^3 + n E_6(\tau)^2$.
This is because $E_4(\tau)$ contains order-$\frac13$ zeros. 
Therefore, in order to have an order-1 zero, we must consider a linear combination of $E_4^3$ and $E_6^2$, which is a modular form of weight 12.
Thus, we find that the Wro\'nskian $W_3$ is proportional to $E_6(\tau)(m E_4(\tau)^3 + n E_6(\tau)^2)$, as it is constant in $z$, which makes it a modular form of weight 18. 

The zeros of the Wro\'nskian are the poles of the coefficients $\phi_r(z,\tau)$, in both $z$ and $\tau$; therefore, we can factor out the weight 18 modular form from the denominator:
\begin{equation}
  \begin{split}
    &E_6(\tau) (m E_4(\tau)^3+n E_6(\tau)^2)\partial^3\mc F\\
    &+ (\alpha_1(\tau) \wp(z|\tau)+\alpha_2(\tau))\partial \mc F + \beta_1(\tau) \wp'(z|\tau) \mc F  = 0,
  \end{split}
\end{equation}
where $\alpha_1$ and $\beta_1$ are also weight 18 modular forms and $\alpha_2$ is a weight 20 modular form.
We can also say that $\beta_2 = 0$ since there exist no odd weight modular forms. 

We can express the weight 18 modular forms 
\begin{equation}
  \begin{split}
    \alpha_1 &= E_6(\tau) (m_\alpha E_4(\tau)^3 + n_\alpha E_6(\tau)^2), \\
    \beta_1 &= E_6(\tau) (m_\beta E_4(\tau)^3 + n_\beta E_6(\tau)^2),
  \end{split}
\end{equation}
without loss of generality.
The weight 20 modular form can be expressed as $\alpha_2 = E_4(\tau)^2(m_\gamma E_4(\tau)^3 + n_\gamma E_6(\tau)^2)$ also without loss of generality.

The differential equation must be satisfied by $\frac1z$, $z$, and $z^3$ in the leading order. 
Imposing this helps us determine the parameters defined above. 
We obtain
\begin{equation}
  \begin{aligned}
    m_\alpha &= 2m_\beta = -3 m, \\
    n_\alpha &= 2 n_\beta = -3n, \\
    m_\gamma &= n_\gamma = 0.
  \end{aligned}
\end{equation}
Therefore, the conformal blocks of the correlator $\ev{\sigma(z,\bar z)\sigma(0, 0)}_{\tau,\bar\tau}$ in the $\mathbb Z_2$ cyclic orbifold of the E$_{8,1}$ WZW model satisfy
\begin{equation}
  \label{eq:E8diffeq}
  \partial^3 \mc F - 3\wp(z|\tau) \partial \mc F - \frac32 \wp'(z|\tau) \mc F= 0.   
\end{equation}

We see that the zeros of the Wro\'nskian $W_3$ are canceled out by the lower rank Wro\'nskians. 
This is likely due to the fact that the dimensions of the rings of modular forms of weights 12, 18, and 20 are all two, and the Wro\'nskian only had derivatives in $z$, therefore not changing the locations of the zeros in $\tau$.

\subsection{Solving the differential equation}

Expressing a doubly periodic differential equation, such as \eqref{eq:E8diffeq}, in terms of the Weierstra{\ss} $\wp$ function can facilitate the identification of its symmetries; however, it becomes less convenient when trying to solve the equation.
An alternative and more practical approach involves using Jacobi elliptic functions to represent the Weierstra{\ss} $\wp$ function \cite{ARSCOTT1964191, Whittaker_Watson_1996}.
To maintain consistency with the previous section, we present the solutions in terms of theta functions. 
Then, one can verify that 
\begin{equation}\label{eq:E8sols}
  \frac{\vartheta_i(z|\tau)}{\vartheta_1(z|\tau)}, \quad i = 2,3,4  
\end{equation}
satisfies the differential equation \eqref{eq:E8diffeq}.
We list the identities necessary to check the equivalence between the theta functions and the elliptic functions in Eq. \eqref{eq:el2th}.
The Weierstra{\ss} $\wp$ function and its derivative in terms of Jacobi theta functions can be found in Eq. \eqref{eq:wptotheta}.
Furthermore, this set of linearly independent functions satisfies the requirements of conformal blocks as functions of $z$ as detailed in Sec. \ref{sec:review}.
Interestingly, Eq. \eqref{eq:E8diffeq} is also applicable to the $\mathrm{SU}(2)_2$ WZW model.
This model is equivalent to a theory of three free Majorana fermions with a central charge $c = \frac{3}{2}$.
In this context, the two-point correlation function is defined by the Szeg\"o kernels, which align with the solutions \eqref{eq:E8sols} \cite{Mathur:1988yx,Fay:1973}.

\subsection{Normalizing the solutions to obtain conformal blocks}

To obtain the conformal blocks, we need to properly normalize the solutions to the differential equation. 
Right now, the solutions \eqref{eq:E8sols}
are not in the appropriate basis to demand the normalization condition, 
\begin{equation}\label{eq:normalizationcondition}
  \lim\limits_{z\to 0} z \mc F_i(z|\tau) = \chi_i(\tau).
\end{equation}
It is clear to see from the behavior of the solutions under the modular $\mc S$ transformation
\begin{equation}
  \begin{pmatrix}
    \frac{\vartheta_2\left(\frac z\tau|-\frac1\tau\right)}{\vartheta_1\left(\frac z\tau|-\frac1\tau\right)}\\
    \frac{\vartheta_3\left(\frac z\tau|-\frac1\tau\right)}{\vartheta_1\left(\frac z\tau|-\frac1\tau\right)}\\
    \frac{\vartheta_4\left(\frac z\tau|-\frac1\tau\right)}{\vartheta_1\left(\frac z\tau|-\frac1\tau\right)}
  \end{pmatrix}
  =
  \begin{pmatrix}
    0&0&i\\
    0&i&0\\
    i&0&0
  \end{pmatrix}
  \begin{pmatrix}
    \frac{\vartheta_2(z|\tau)}{\vartheta_1(z|\tau)}\\
    \frac{\vartheta_3(z|\tau)}{\vartheta_1(z|\tau)}\\
    \frac{\vartheta_4(z|\tau)}{\vartheta_1(z|\tau)}
  \end{pmatrix}
\end{equation}
that this basis is not the same basis as that of the characters. 
However, we note the observation that the following basis of the Klein-$j$ functions making up the characters transforms in the same way as the solutions under a modular $\mc S$ transformation:
\begin{equation}
  \mc S
  \begin{pmatrix}
    j\left(\frac\tau2\right)^{\frac13}\\    j(\tau)^{\frac23}\\j(2\tau)^{\frac13}\\
  \end{pmatrix}
  =
  \begin{pmatrix}
    j\left(-\frac1{2\tau}\right)^{\frac13}\\    j\left(-\frac1\tau\right)^{\frac23}\\j\left(-\frac2\tau\right)^{\frac13}\\
  \end{pmatrix}
  =\begin{pmatrix}
    0&0&1\\0 &1&0\\ 1&0&0\\
  \end{pmatrix}
  \begin{pmatrix}
    j\left(\frac\tau2\right)^{\frac13}\\    j(\tau)^{\frac23}\\j(2\tau)^{\frac13}\\
  \end{pmatrix}.
\end{equation}
This basis is ideal to normalize the solutions such that the final blocks transform appropriately under the modular $\mc S$ transformation, since the solutions only exchange and do not go into nontrivial linear combinations of themselves. 
In other words, a diagonal or antidiagonal basis is an ideal choice of basis to perform the normalization.
Note that the choice of basis is not unique, since both $(j\left(\frac\tau2\right)^{\frac13},j(\tau)^{\frac23}, j(2\tau)^{\frac13})^{\mathsf T}$ and $(j(2\tau)^{\frac13},j(\tau)^{\frac23}, j\left(\frac\tau2\right)^{\frac13})^{\mathsf T}$ have the same $\mc S$ matrix.
To check which of the choices is correct, we can perform a $q$-expansion of the blocks in the correct basis and use the choice which yields the correct behavior. 

Now, let us find the normalizations, taking into account the ambiguity in the choice of basis. 
Using the identity \eqref{eq:theta1lim} and the normalization condition \eqref{eq:normalizationcondition} we find that the solutions normalized in the current basis yield the following normalizations:
\begin{equation}
  \begin{split}
    \lim_{z\to 0} z N_0(\tau) \frac{\vartheta_2(z|\tau)}{\vartheta_1(z|\tau)}
    &= j\left(\tfrac\tau2\right)^{\frac13} \Rightarrow N_0(\tau)
    = \frac{2\pi j\left(\frac\tau2\right)^{\frac13}\eta(\tau)^3}{\vartheta_2(\tau)},\\
    \lim_{z\to 0} z N_1(\tau) \frac{\vartheta_3(z|\tau)}{\vartheta_1(z|\tau)}
    &= j(\tau)^{\frac23} 
    \Rightarrow N_1(\tau) = \frac{2\pi j\left(\tau\right)^{\frac23}\eta(\tau)^3}{\vartheta_3(\tau)},\\
    \lim_{z\to 0} z N_2(\tau) \frac{\vartheta_4(z|\tau)}{\vartheta_1(z|\tau)}
    &= j(2\tau)^{\frac13}
    \Rightarrow N_2(\tau) = \frac{2\pi j\left(2\tau\right)^{\frac13}\eta(\tau)^3}{\vartheta_4(\tau)},\\
  \end{split}
\end{equation}
where $N_i$ is the $\tau$-dependent normalization.
The other choice of basis amounts to switching $\vartheta_2$ and $\vartheta_4$. 

Let us note the change of basis matrix from the Klein-$j$ invariants to characters
\begin{equation}
  \begin{split}
    \begin{pmatrix}  \chi_0(\tau)\\\chi_1(\tau) \\ \chi_2(\tau)  \end{pmatrix}
    &= \begin{pmatrix}
      \frac12(j(\tau)^{\frac23} + j(2\tau)^{\frac13})\\
      \frac12(j(\tau)^{\frac23} - j(2\tau)^{\frac13}) \\
      \frac12(2j\left(\frac\tau2\right)^{\frac13} - j(\tau)^{\frac23}+ j(2\tau)^{\frac13})\\
    \end{pmatrix}\\
    &=
    \frac12
    \begin{pmatrix}
      0& 1& 1\\
      0& 1&-1\\
      2&-1& 1\\
    \end{pmatrix}
    \begin{pmatrix}
      j\left(\frac\tau2\right)^{\frac13}\\    j(\tau)^{\frac23}\\j(2\tau)^{\frac13}\\
    \end{pmatrix}.
  \end{split}
\end{equation}

Using the same change of basis matrix on the normalized solutions, we obtain the conformal blocks in the OPE channel, where the condition \eqref{eq:normalizationcondition}, is satisfied.
Therefore, the conformal blocks read
\begin{widetext}
  \begin{equation}\label{eq:confblocks}
    \begin{split}
      \mc F_0(z|\tau)
      &= \frac{\pi\eta(\tau)^3}{\vartheta_1(z|\tau)}
      \left(\frac{j(\tau)^{\frac23} \vartheta_ 3(z|\tau)}{\vartheta_3(\tau)}
        +\frac{j(2\tau)^{\frac13} \vartheta_ 4(z|\tau)}{\vartheta_ 4(\tau)}\right),\\
      \mc F_1(z|\tau)
      &= \frac{\pi\eta(\tau)^3}{\vartheta_1(z|\tau)}
      \left(\frac{j(\tau)^{\frac23} \vartheta_ 3(z|\tau)}{\vartheta_3(\tau)}
        -\frac{j(2\tau)^{\frac13} \vartheta_ 4(z|\tau)}{\vartheta_ 4(\tau)}\right),\\
      \mc F_2(z|\tau)
      &=\frac{\pi\eta(\tau)^3}{\vartheta_1(z|\tau)}\left(
        \frac{j(2\tau)^{\frac13} \vartheta_4(z,\tau)}{\vartheta_4(\tau)}
        +\frac{2j\left(\frac\tau2\right)^{\frac13} \vartheta_2(z|\tau)}{\vartheta_2(\tau)}
        -\frac{j(\tau)^{\frac23} \vartheta_3(z|\tau)}{\vartheta_3(\tau)}\right). 
    \end{split}
  \end{equation}
\end{widetext}
One can easily verify that, with this choice of normalization, the modular $\mc S$ matrix $M^{(\mc S)}$ is proportional to the modular $\mc S$ matrix for the characters.
The conformal blocks are plotted in Appendix \ref{app:figures}.

Expanding the conformal blocks \eqref{eq:confblocks} in $q$, we find
\begin{equation}
  \begin{split}
    \frac1\pi \mc F_0(z|\tau)
    &= \frac{q^{-\frac23}}{\sin(\pi z)} + \left(\frac{248}{\sin(\pi z)} + 4\sin(\pi z) \right)q^{\frac13} \\
    &\qquad  - 992 \sin(\pi z) q^{\frac56} + \mc O(q^{\frac43}),\\
    \frac1\pi \mc F_1(z|\tau)
    &= -4\sin(\pi z)q^{-\frac16} + \left(\frac{248}{\sin(\pi z)}\right)q^{\frac13}\\
    &\qquad  - 8(125 + \cos(2\pi z)) \sin(\pi z) q^{\frac56} + \mc O(q^{\frac43}),\\
    \frac1\pi \mc F_2(z|\tau)
    &= (4\sin(\pi z) + \cot(\pi z) )q^{-\frac16} -248\tan(\frac{\pi z}2)q^{\frac13} \\
    &\qquad +2(2061\cos(\pi z)-248\cos(2\pi z)\\
    &\qquad +\cos(3\pi z)-\cos(4\pi z)+249) q^{\frac56} + \mc O(q^{\frac43}),
  \end{split}
\end{equation}
where we see the conformal blocks' leading $q$ behaviors are as predicted.
The other basis with $\vartheta_2$ and $\vartheta_4$ switched does not have the correct leading $q$ behavior, ruling it out.
Similarly, in the $z\to 0$ limit, all blocks have the appropriate powers of $q$ in the leading order, which match the behavior with the corresponding character.

Now that we have the correct choice for the conformal blocks, let us compute their behavior under the periodicities.
Using the periodicity properties of the Jacobi theta functions we can calculate the transformation matrices for the solution vector $N_i(\tau)\frac{\vartheta_{i+1}(z|\tau)}{\vartheta_1(z|\tau)}$, and then apply the change of basis matrix used above to calculate the monodromy matrices $M^{(1)}$ and $M^{(\tau)}$.
We find
  \begin{equation} M^{(1)} = 
    \begin{pmatrix}
      -1 & 0 & 0 \\
      0 & -1 & 0 \\
      0 & 2 & 1 \\
    \end{pmatrix}, \;
    M^{(\tau)} = 
    \begin{pmatrix}
      0 & -1 & 0 \\
      -1 & 0 & 0 \\
      1 & -1 & -1 \\
    \end{pmatrix}.
  \end{equation}
Both $M^{(1)}$ and $M^{(\tau)}$ square to one, which implies that the conformal blocks are doubly-periodic with periods 2 and $2\tau$.
We plot the conformal blocks in Fig. \ref{fig:conformalblocks}, where it is easily visible that the conformal blocks are (2,2$\tau$)-periodic along the lattice vectors $1$ and $\tau$.
We also find that $M^{(1)}$ and $M^{(\tau)}$ commute, so the correlator at $z+1+\tau$ is the same as $z+\tau+1$, as expected.

Finally, we can express the full correlator:
\begin{equation}
  \label{eq:fullcorrelator}
  \begin{split}
    &\ev{\sigma(z_1,\bar z_1)\sigma(z_2,\bar z_2)}_{\tau,\bar\tau} \\
    &= \frac{|\mc F_0(z_{12}|\tau)|^2 + 2|\mc F_1(z_{12}|\tau)|^2 + |\mc F_2(z_{12}|\tau)|^2}{Z_{\text{orb}}(\tau,\bar \tau)} ,
  \end{split}
\end{equation}
where $Z_{\rm orb}$ is given in \eqref{eq:Zorb}, 
which in the $\tau,\bar\tau\to i\infty$ limit yields
\begin{equation}
  \lim_{\substack{\tau\to i\infty\\\bar\tau\to i\infty}}
  \ev{\sigma(z_1,\bar z_1)\sigma(z_2,\bar z_2)}_{\tau,\bar\tau} = \frac{\pi^2}{|\sin(\pi z_{12})|^2}, 
\end{equation}
reproducing the behavior of a two-point function of primary fields with conformal weight $\frac12$ on the cylinder of unit circumference.

The first few subleading terms in the $\tau\to i\infty, \bar\tau \to i\infty$ limit are displayed in the Appendix \ref{app:exps} in Eq. \eqref{eq:exp0}. 
Note that the singularity at $z=\bar z = 0$ can be completely factored out from the $q$-expansion. 

To study the periodicity properties of the full correlator, and therefore the R\'enyi entropy, we can use the properties of the monodromy matrices $M^{(1)}$ and $M^{(\tau)}$. 
Since they square to 1, we can immediately say that the correlator has periods 2 and $2\tau$.
To see if the correlator is one-periodic along both cycles, we must see if the monodromy matrices preserve the degeneracy matrix.
The degeneracy matrix is $D=\mathrm{diag}(1,2,1)$, so we have
\begin{equation}
  \begin{aligned}
    M^{(1)\mathsf T} D M^{(1)}
    &=
      \begin{pmatrix}
        1 & 0 & 0 \\
        0 & 6 & 2 \\
        0 & 2 & 1 \\
      \end{pmatrix} \neq D, \\
    M^{(\tau)\mathsf T} D M^{(\tau)}
    &=
      \begin{pmatrix}
        3 & -1 & -1 \\
        -1 & 2 & 1 \\
        -1 & 1 & 1 \\
      \end{pmatrix}
      \neq D.
  \end{aligned}
\end{equation}

Therefore, we see that the correlator is not one-periodic along either cycle of the torus as the degeneracy matrix is not preserved on translations along either cycle.
Similar periodic properties were found in \cite{Atick:1986ns}, in which correlators of spin operators are calculated on the torus using Green's function techniques.
Since the spin operators are fermionic, as the spin operator goes around a cycle of the torus, it changes the twisted sector/fermionic boundary condition in which the correlator is calculated.
This suggests that the twist operator behaves similarly.
As the twist operator goes around a cycle of the torus, it changes the twisted sector, just as translating a spin operator around the torus changes the fermionic boundary conditions.

Studying the poles of the torus, we find that the poles are only at the lattice points $z\in m\tau + n$, where $m,n \in \mathbb Z$.
This is expected as the lattice points are also the coincident limit.
The residue of the pole when $m$ is even is $\pi^2$, whereas the residue of the pole when $m$ is odd is $3\pi^2$.
This can be seen in the $q$-expansions when $z\to z+1$  and $z\to z+\tau$, which are listed in Appendix \ref{app:exps}, as well as in Fig. \ref{fig:correlator}.
Physically, these expansions can be interpreted as taking the decompactification limit $\tau\to i\infty$ after going around the torus once.
This probes the theory in two different twisted sectors that are not connected perturbatively, similar to expanding a theory around two different saddle points.

Another interpretation of the divergences of the correlator can be interpreted as the singular limit of the genus 2 manifold.
For example, the cylinder can be thought of as the singular limit of the torus when one takes the modulus to infinity and the partition function diverges as $q^{-\frac{c}{24}}$.
The moduli of the replicated torus are $\tau$ and $z$, where $\tau$ is the modulus of the individual tori and $z$ is the interval along which the tori are glued.
We find that $z=\left\{0,1,1+\tau\right\}$ are singular points in the moduli space because the twist correlator can be interpreted as the partition function of the CFT on the genus 2 Riemann surface, and is divergent at these points.
This suggests that the manifold created by gluing two tori along an interval which wraps around a cycle of the torus is singular.

\begin{figure}
  \centering
  \includegraphics[height=70mm]{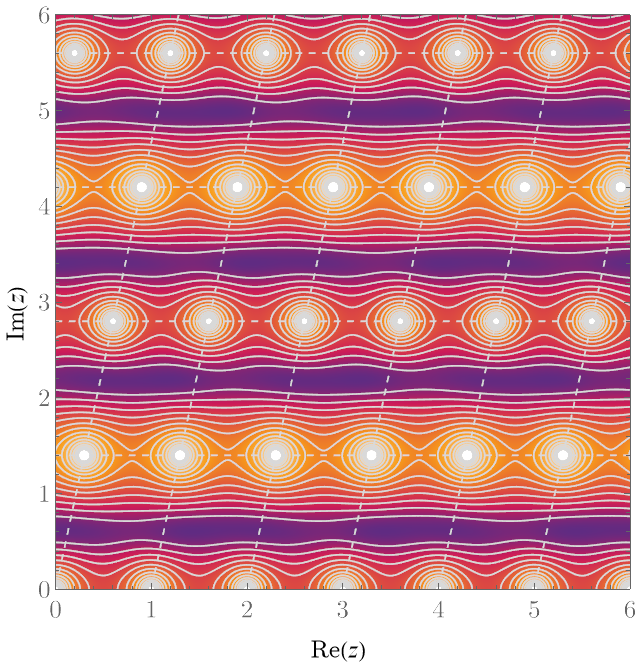}
  \includegraphics[width=70mm]{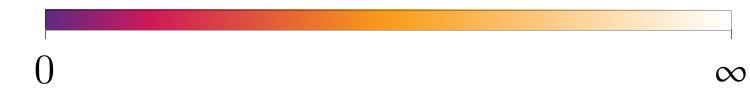}
  \caption{\justifying
    Two-dimensional visualization of the two-point function \eqref{eq:fullcorrelator} where $\tau = 0.3 + 1.4 i$, $\bar \tau = - \tau^*$, $\bar z=z^*$, and $z$ takes values between 0 and $6+6i$.
    The color gradient represents the magnitude, and poles are specifically highlighted in white.
    The dashed lines indicate the lattice $\mathbb Z+\tau \mathbb Z$ and the contours form loci of equal values.}
  \label{fig:correlator}
\end{figure}

As discussed in the Introduction, the R\'enyi entropy of a subregion $A$ between $(0,0)$ and $(z,\bar z)$ can be expressed in terms of the correlator of twist operators,
\begin{equation}
  \begin{split}
    S_N(A) &= \frac1{1-N}\ln \Tr \rho_A^N  \\
    &= \frac{1}{1-N}\ln\ev{\sigma(z,\bar z)\bar \sigma(0,0)}.
    \end{split}
\end{equation}
Let us expand the second R\'enyi entropy on the torus for a single interval for the E$_{8,1}$ WZW model in the decompactified limit,
\begin{widetext}
  \begin{equation}
    \begin{split}
      S_2(A) &= 2\ln\left|\frac{\sin(\pi z)}\pi\right| +
               \big(1
               -4\sin^2(\pi z)\left(12\sin^2(\pi\bar z)+\cos(\pi\bar z)\right)
               -\cos(\pi z)\left(4\sin^2(\pi\bar z)+\cos(\pi\bar z)\right)\big)\sqrt{q\bar q}\\
             &\qquad
               -4\sin^2(\pi z) q
               -248 \left((\cos (\pi  z)-1) \cos (\pi  \bar z)+4 (\cos (\pi  z)-3) \sin ^2(\pi  \bar z)\right)q\sqrt{\bar q}\\
             &\qquad
               - 4\sin^2(\pi\bar z)\bar q
               -248 \left(\cos (\pi  z) (\cos (\pi  \bar z)-1)+4 \sin ^2(\pi  z) (\cos (\pi  \bar z)-3)\right)\bar q\sqrt q\\
             &\qquad
               \frac18
               \bigg(4 (-123000 \cos (\pi  z)-47 \cos (2 \pi  z)-8 \cos (3 \pi  z)
               +12 \cos (4 \pi  z)+123045) \cos (\pi  \bar z)\\
             &\qquad
               +8 \cos (4 \pi  \bar z) \left(12 \sin ^2(\pi  z)+\cos (\pi  z)\right)^2
               -8 \cos (3 \pi  \bar z) \left(4 \sin ^2(\pi  z)+\cos (\pi  z)\right) \left(12 \sin ^2(\pi  z)+\cos (\pi  z)\right)\\
             &\qquad
               -2 \cos (2 \pi  \bar z) \left(44 \sin ^2(\pi  z)+3 \cos (\pi  z)\right) \left(52 \sin ^2(\pi  z)+5 \cos (\pi  z)\right)
               +492180 \cos (\pi  z)-1731 \cos (2 \pi  z)\\
             &\qquad
               -148 \cos (3 \pi  z)
               +436 \cos (4 \pi  z)-490715\bigg)
               q\bar q + \mc O(q^{\frac32}) +\mc O(\bar q^{\frac32}).
    \end{split}
  \end{equation}
\end{widetext}

Since the divergent term at $z=0$ factored out from the correlator, we find that the finite $q$-corrections to the second R\'enyi entropy are all UV finite.
In other words, the universal divergent piece is not $q$-dependent, in the low-temperature limit.
This suggests that this structure extends to more than this specific CFT.
The UV-finite corrections may encode interesting correlations between the operators of the two subregions, and may provide a way to compute the spectrum of operators such as the modular Hamiltonian of Tomita-Takesaki theory.

Furthermore, the periodicity properties of the R\'enyi entropy are inherited from the periodicities of the twist two-point correlator.
This means that the R\'enyi entropy is also two-periodic along both cycles of the torus.
In the case of free fermions, this was encountered in \cite{Mukhi:2017rex, Mukhi:2018qub}, where it was found that the $N$th R\'enyi entropy is $N$-periodic along each cycle, as the Riemann surface is the $N$-fold cover of the torus, and is genus $N$.
Therefore, another interpretation of our result that the twist two-point correlator is two-periodic is that the correlator is actually defined on a genus 2 manifold rather than on a genus 1 torus.
This should be expected, since the twist two-point correlator is the partition function of the theory on the replicated surface, where the moduli are the modular parameters of the replicated torus, and the separation of the twist operators that corresponds to the length of the interval along which the two tori are glued.

\section{Discussion}
\label{sec:discussion}

In this paper, we show how one can use the Wro\'nskian method of \cite{Mathur:1988rx} to calculate R\'enyi entropies for rational CFTs on the torus.
The calculation of torus R\'enyi entropies have been limited to free theories, since the techniques have required explicit computation of objects like propagators and resolvent kernels on the torus, which is only tractable for free theories.
However, using properties of elliptic functions and modular forms, it is possible to use the Wro\'nskian method to constrain differential equations on the torus, which allows one to compute correlation functions for more complicated CFTs.

To demonstrate the power of this procedure, we calculate the second R\'enyi entropy of a single interval for the $\rm E_{8,1}$ WZW model on a torus. 
First, we have to find the primaries that arise from the cyclic orbifolding procedure that the replica trick entails. 
To do so, we compute the $\mathbb Z_2$ cyclic orbifold partition function of all single-character CFTs and find that the orbifold results in a four-character CFT.
However, for the $\rm E_{8,1}$ WZW model, two of the four characters become degenerate, and thus yield a three-character CFT.
We also identify that all $\mathbb Z_2$ cyclic orbifolds of meromorphic CFTs are realizations of the toric code modular tensor category, due to its vanishing topological central charge at rank 4. 

Following this, we construct and solve the differential equation for the conformal blocks of the twist two-point function on the torus. 
Normalizing the conformal blocks appropriately and combining them gives us the expression for the two-point function. 
We see that the leading behavior in the decompactification limit behaves as expected for a CFT on the cylinder.
Furthermore, we see that the leading singularity is shared with all higher-order corrections in the nome $q$.

Using the twist two-point function, we compute the second R\'enyi entropy. 
Since the divergent term factors out, we find that on subtracting the universal divergent piece from the $q$-expansion of the R\'enyi entropy, the result is completely finite even in the coincident limit. 
This suggests that the correlations between operators of the two subregions have a rich structure beyond the infinite entanglement suggested by the UV-divergent piece, which is cutoff dependent.

We also find that the two-point correlator, and in turn the R\'enyi entropy, are not periodic along each cycle of the torus, but two-periodic.
One way to interpret this is that as the twist operator is taken around a cycle of the torus, it changes the twisted sector in which the computation takes place.
Since we are considering a $\mathbb Z_2$ orbifold, going around the cycle a second time recovers the original twisted sector.
Another interpretation is that the Riemann surface is not genus 1, but genus 2, since the theory is on a replicated torus.
Therefore, the geometry of the surface shows up in the periodicity properties of the correlator. 

It would be interesting to understand the implications of this result in terms of operator algebras. 
It is clear that the divergence in the R\'enyi entropy is a result of the fact that quantum field theories without a UV cutoff have a von Neumann factor of type III$_1$.
However, it would be interesting to understand what the implications of the nonuniversal finite terms for the operator algebra are, as these terms depend on the geometry of the manifold the field theory is defined.

New directions include the computation of other quantum information measures, such as mutual R\'enyi entropy, distance entropy, and other related measures.
The second R\'enyi entropy encodes the probability that two random variables from identical distributions are equal, and is also called the collision entropy.
Extending this procedure to a larger number of replicas should allow us to compute quantities like the min entropy, which requires taking the number of replicas to infinity.
Having the R\'enyi entropies with a different number of replicas should allow one to numerically interpolate the behavior to a single replica to compute the von Neumann entropy. 

Another interesting avenue of exploration would be to extend our results to a large central charge and explore the holographic limit, by taking $k\to \infty$.
Finding how the conformal blocks rearrange at a large central charge to construct boundary to boundary propagators of fields in the bulk would shed light on how the AdS/CFT correspondence works and give us insight into how to go away from the semiclassical limit.

Another immediate future direction would be to extend this procedure to RCFTs with more than one character.
The differential equations would be of higher order, since more characters imply a larger number of conformal blocks, but may still be solvable numerically.
Furthermore, it would be very interesting to understand how integrable deformations such as the $T\overline T$ deformation affects the R\'enyi entropy nonperturbatively, to extend this procedure to field theories without conformal symmetry.

\section*{AACKNOWLEDGMENTS}
\label{sec:acknowledgements}

The authors thank Fri{\dh}rik Freyr Gautason, Valentina Giangreco M. Puletti, Matthias Harksen, Victoria Martin, Viraj Merulia, Vyshnav Mohan, Sunil Mukhi, and L\'arus Thorlacius, for very useful and insightful conversations. 
We also thank the anonymous referee for the valuable comments.
This work was supported in part by the University of Iceland Research Fund and by the Icelandic Research Fund under grant 228952-053.

\onecolumngrid

\appendix

\section{THE TORUS AND THE MODULAR GROUP}
The complex torus can be realized as the quotient space $\mathbb C/\Lambda$, where $\Lambda = \omega_1 \mathbb Z + \omega_2 \mathbb Z$ is a lattice on the complex plane, whose lattice vectors $\omega_1$ and $\omega_2$ corresponded to the periods of the torus.
The lattice $\Lambda$ can be equivalently described by choosing a new set of lattice vectors $\omega_1'$ and $\omega_2'$, such that $\omega_1 \mathbb Z + \omega_2 \mathbb Z = \omega_1' \mathbb Z + \omega_2'\mathbb Z$, which implies
\begin{equation}\label{eq:newlat}
  \begin{pmatrix}
    \omega_2'\\\omega_1'
  \end{pmatrix}
  = \begin{pmatrix}
    a & b\\ c & d
  \end{pmatrix}
  \begin{pmatrix}
    \omega_2\\\omega_1
  \end{pmatrix}\,,
  \quad
  \begin{pmatrix}
    a & b\\ c & d
  \end{pmatrix} \in \mathrm{SL}_2(\mathbb Z).
\end{equation}
Since we are considering the geometry up to a scaling we can factor out $\omega_1$ and we utilize $\Lambda = \mathbb Z + \tau \mathbb Z$ where $\tau=\frac{\omega_2}{\omega_1}$ is called the modular parameter.
Without loss of generality $\tau$ belongs to the upper half-plane $\mathbb H = \left\{\tau\in\mathbb C \mid \mathrm{Im}(\tau)>0\right\}$.
Since $-\mathds 1$ acts trivially on $\mathbb H$, the symmetry group is $\mathrm{PSL}_2(\mathbb Z) = \mathrm{SL}_2(\mathbb Z)/\pm\mathds1$, known as the modular group.

Using \eqref{eq:newlat}, we can calculate how the modulus transforms under a $\mathrm{PSL}_2(\mathbb Z)$ transform,
\begin{equation}
  \gamma\tau \equiv \frac{\omega_2'}{\omega_1'} = \frac{a\tau +b}{c\tau+d}\quad \forall \gamma\in \mathrm{PSL}_2(\mathbb Z).
\end{equation}
Therefore, the space of inequivalent lattices is given by the quotient $\mathrm{PSL}_2(\mathbb Z)\backslash \mathbb H$.
The canonical representative of the quotient is called the fundamental domain, and is conventionally defined by 
\begin{equation}
  \mathbb F = \left\{\tau\in\mathbb H \;\bigg|\; |\tau|\geq1,\;\left|\mathrm{Re}(\tau )\right|\leq\frac12\right\}.
\end{equation}
The fundamental domain is highlighted in Fig. \ref{fig:FundDom}, and is a hyperbolic triangle whose corners are $e^{i\frac\pi3}, e^{i\frac{2\pi}3},i\infty$.
The first two are called elliptic points, and the third is called the cusp.
A third elliptic point is at $\tau = i$, which is not a corner of the hyperbolic triangle.
Therefore, the action of $\mathrm{PSL}_2(\mathbb Z)$ generates a \emph{tesselation} of the hyperbolic plane.
The generators of the modular group are
\begin{equation}
  \mathcal S \colon \tau \mapsto -\frac1\tau,\quad \text{and}\quad \mathcal T \colon \tau \mapsto \tau+1.
\end{equation}

\begin{figure}
  \centering
  \begin{tikzpicture}[scale=1.8]
    \node at (2.3,2.6) {$\mathbb H$};
    \shade[bottom color=red,top color=white] (-0.5,0.866025) rectangle (0.5,3);
    \filldraw[white] (1,0) arc(0:180:1);
    \draw[->, thick] (0,0) -- (0,3) node [anchor=west] {$\Im\tau$};
    \foreach \x in {-2,...,3} \draw[->,gray] (\x-1/2,0) -- (\x-1/2,3);
    \foreach \y in {1,3,...,21} \foreach \x in {-3,...,2} \draw[thin,gray] (\x+2/\y,0) arc(0:180:1/\y);
    \foreach \y in {1,3,...,21} \foreach \x in {-3,...,2} \draw[thin,gray] (\x+1,0) arc(0:180:1/\y);
    \foreach \y in {8,10,...,20} \foreach \x in {-2,...,3} \draw[thin,gray] (\x- 1/2 + 2/\y,0) arc(0:180:1/\y);
    \foreach \y in {8,10,...,20} \foreach \x in {-2,...,3} \draw[thin,gray] (\x- 1/2,0) arc(0:180:1/\y);
    \filldraw[white] (-4.1,4)rectangle (-3,0);
    \filldraw[white] (3,4)rectangle (4.1,0);
    \draw[<->, thick] (-3.1,0) -- (3.1,0) node [anchor=north] {$\Re \tau$};
    \filldraw (60:1) circle (0.2mm) node [anchor=south west] {$\rho$};
    \filldraw (120:1) circle (0.2mm)node [anchor=south west] {$\rho^2$};
    \filldraw (90:1) circle (0.2mm)node [anchor=south west] {$i$};
    \filldraw (90:3.2) circle (0.2mm)node [anchor=south west] {$i\infty$};
  \end{tikzpicture}
  \caption{\justifying The fundamental domain $\mathbb F = \mathrm{PSL}_2(\mathbb Z)\backslash \mathbb H$ highlighted in red.
    The elliptic points $\tau=e^{\frac{2i\pi}3}\equiv\rho^2$, $\tau=i$ and $\tau=e^{\frac{i\pi}3}\equiv\rho$ and the cusp at $\tau=i\infty$.
    The fundamental domain can be mapped to any other subregion, shown as enclosed hyperbolic triangles, by a $\mathrm{PSL}_2(\mathbb Z)$ transformation.}
  \label{fig:FundDom}
\end{figure}

Let $z$ be the coordinate on the torus, such that the periodicities are $z \sim  z+1 \sim z+\tau$.
The coordinate transforms under a $\mathrm{PSL}_2(\mathbb Z)$ transformation due to its periodicity $z\sim z+\tau$,
\begin{equation}
  \begin{split}
    \gamma z \sim \gamma z + \gamma \tau 
    &= \gamma z + \frac{a\tau+b}{c\tau+d} \\
    \implies (c\tau+d)\gamma z &\sim (c\tau + d) \gamma z + a \tau + b.
  \end{split}
\end{equation}
Note that $(c\tau + d)\gamma z$ has the same periodicities as $z$, so we conclude 
\begin{equation}
  \gamma z = \frac{z}{c\tau+d}.
\end{equation}
This transformation will be useful when computing the modular transformations of functions on the torus, which depend on both the coordinate $z$ and the modulus $\tau$.

\section{MODULAR FORMS}
\label{app:modularforms}
Modular forms of weight $k$ are holomorphic functions of $\tau$, which transform covariantly under the modular transformations as follows,
\begin{equation}\label{eq:mform}
  f\left(\gamma\tau\right) = \left(c\tau + d\right)^k f(\tau).
\end{equation}
We saw that the coordinate on the torus $z$ transforms as a modular form of weight $-1$.
Other canonical examples of modular forms are the Eisenstein series $G_{2k}$.
These are modular forms of weight $2k$ for $k\geq2$, which are defined by
\begin{equation}
  G_{2k}(\tau)  \equiv \sum\limits_{\substack{m,n\in \mathbb Z\\\{m,n\}\neq\{0,0\}}} \frac{1}{\left(m\tau+n\right)^{2k}} \equiv 2\zeta(2k) E_{2k}(\tau),
\end{equation}
where $E_{2k}$ are the normalized Eisenstein series and $\zeta$ is the Riemann zeta function.
The Fourier expansions for the first two can be calculated using the following series: 
\begin{align}
  E_{4}(\tau) = 1 + 240\sum\limits_{n\in\mathbb Z^+}\frac{n^3 q^n}{1 - q^n} ,\quad  
  E_{6}(\tau) = 1 - 504\sum\limits_{n\in\mathbb Z^+}\frac{n^5 q^n}{1 - q^n},
\end{align}
where $q = e^{2\pi i\tau}$.

It turns out that all modular forms can be expressed as linear combinations of products of $E_4$ and $E_6$,  
\begin{equation}
  f_{k}(\tau) = \sum_{i} n_iE^{a_i}_4(\tau)E_6^{b_i}(\tau) 
\end{equation}
where the sum is over all integer pairs $(a_i,b_i)$ such that $4a_i+6b_i=k$, since the product of two modular forms of weight $k_1$ and $k_2$ create a new modular form of weight $k_1+k_2$, following from the definition of modular forms.
For example, modular forms of weight 12 can be formulated as a linear combination of $E_4(\tau)^3$ and $E_6(\tau)^2$.
Similarly, those of weight 18 can be constructed as a linear combination of $E_4(\tau)^2E_6(\tau)$ and $E_6(\tau)^3$, and those of weight 20 as a linear combination of $E_4(\tau)^5$ and $E_4(\tau)^2E_6(\tau)^2$.

Note that the fundamental domain $\mathbb F$ contains special points at the corners and edges of the hyperbolic triangle, which contain conical defects due to the action of the quotient.
In other words, the monodromy around the zeros and poles at these points will not be a full $2\pi$.
Specifically, the monodromy at the elliptic points $\tau=i$ and $\tau=e^{i\frac\pi3}=e^{i\frac{2\pi}3}$ have a conical defect of $\pi$ and $\frac{2\pi}3$ respectively. 
Therefore, as a consequence of the Riemann-Roch theorem, a modular form $f$ of weight $k$ obeys the valence formula, and is given by,
\begin{equation}\label{eq:valence}
  \nu_f(i\infty) + \frac12 \nu_f(i) + \frac13 \nu_f(e^{i\frac\pi3})
  + \sum_{\substack{p\in \mathrm{PSL}_2(\mathbb Z)\backslash \mathbb H\\p\neq \{i\infty, i, e^{i\frac\pi3}\}}} \nu_f(p) = \frac{k}{12},
\end{equation}
where $\nu_f(p)$ denotes the order of the zero or pole of $f$ at the point $p$. 

Another canonical modular form is the Dedekind eta function, which is a modular form of weight $\tfrac12$, and can be expressed in terms of the Eisenstein series
\begin{equation}
  \eta(\tau) = q^{\frac{1}{24}}\prod\limits_{n=1}^{\infty}(1-q^n) = \left(\frac{E_4(\tau)^3 - E_6(\tau)^2}{12^3}\right)^\frac{1}{24} \equiv \Delta(\tau)^\frac1{24},
\end{equation}
where $\Delta(\tau)$ is known as the modular discriminant, and is a cusp form, a modular form with a zero at the cusp $\tau=i\infty$.

Additionally, the Klein-$j$ invariant is a function that is invariant under the modular transformation \eqref{eq:mform}, i.e., $j(\gamma \tau) = j(\tau)$.
It is defined by the quotient
\begin{equation}\label{eq:jdef1}
  j(\tau) = \frac{{E_4(\tau)}^3}{\Delta(\tau)} = 1728\frac{E_4(\tau)^3}{{E_4(\tau)}^3-{E_6(\tau)}^2}.
\end{equation}
and has a pole at the cusp $\tau=i\infty$.

\section{JACOBI THETA AND ELLIPTIC FUNCTIONS}
\label{app:elliptic-functions}

Here we list our definitions and conventions for the Jacobi theta functions, in terms of $q = e^{2\pi i \tau}, y = e^{2\pi i z}$, in both sum and product form \cite{Polchinski:1998rq},
\begin{equation} \label{eq:ThetaDefs}
  \begin{split}
    \vartheta_1(z|\tau)
    &= -\sum_{n\in\mathbb Z-\frac12}(-1)^n q^{n^2/2}y^n
      = -iq^{\frac18}y^{\frac12}\prod_{m=1}^\infty (1-q^m)(1-y q^m)(1-y^{-1}q^{m-1}), \\
    \vartheta_2(z|\tau)
    &= \sum_{n\in\mathbb Z-\frac12} q^{n^2/2}y^n
      = q^{\frac18}y^{\frac12}\prod_{m=1}^\infty (1-q^m)(1+yq^m)(1+y^{-1}q^{m-1}),\\
    \vartheta_3(z|\tau) 
    &= \sum_{n\in\mathbb Z}q^{n^2/2}y^n
      = \prod_{m=1}^\infty (1-q^m)(1+y q^{m-\frac12})(1+y^{-1}q^{m-\frac12}),\\
    \vartheta_4(z|\tau)
    &= \sum_{n\in\mathbb Z} (-1)^n q^{n^2/2}y^n
      = \prod_{m=1}^\infty (1-q^m)(1-y q^{m-\frac12})(1-y^{-1}q^{m-\frac12}).
  \end{split}
\end{equation}
When $z = 0$, we drop the argument, i.e, $\vartheta_i(0|\tau) = \vartheta_i(\tau)$.

The theta functions are themselves not elliptic (i.e., doubly periodic), but have the following (quasi)periodic properties:
\begin{equation}\label{eq:JthTransfz}
  \begin{aligned}
    \vartheta_{1,2}(z+1|\tau) &= -\vartheta_{1,2}(z|\tau), &
                                                             \vartheta_{2,3}(z+\tau|\tau) &= y^{-1}q^{-\frac12}\vartheta_{2,3}(z|\tau), \\
    \vartheta_{3,4}(z+1|\tau) &= \vartheta_{3,4}(z|\tau), & 
                                                            \vartheta_{1,4}(z+\tau|\tau) &= -y^{-1}q^{-\frac12}\vartheta_{1,4}(z|\tau).
  \end{aligned} 
\end{equation}
The theta functions can be expressed as translations of the spatial coordinate $z$ of the other theta functions,
\begin{equation}
  \label{eq:ThetaRelns}
  \begin{split}
    \vartheta_2 (z|\tau) &= \vartheta_1\left(\left.z+\tfrac12\right|\tau\right),\\
    \vartheta_3 (z|\tau) &= \vartheta_4\left(\left.z+\tfrac12\right|\tau\right),\\
    \vartheta_2 (z|\tau) &= y^{\frac12}q^{\frac18}\vartheta_3\left(\left.z + \tfrac\tau2\right.|\tau\right).\\
  \end{split}
\end{equation}
The theta functions have no poles and have only a single zero in the fundamental domain of $\mathbb C/ (\mathbb Z + \tau \mathbb Z)$:
\begin{equation} 
  \vartheta_1(0|\tau) = \vartheta_2\left(\tfrac12\big|\tau\right) = \vartheta_3\left(\tfrac{1+\tau}2\big|\tau\right) = \vartheta_4\left(\tfrac\tau2\big|\tau\right) = 0. 
\end{equation}
Only $\vartheta_1$ is an odd function of $z$, while the rest are even.

The modular properties of the theta functions are
\begin{equation}\label{eq:TtransfTheta}
  \begin{split}
    \vartheta_{1,2}(z|\tau+1) &= e^{i\pi/4} \vartheta_{1,2}(z|\tau) ,\quad  \vartheta_{3,4}(z|\tau+1) = \vartheta_{4,3}(z|\tau) ,\\
    \vartheta_1\left(\tfrac z\tau \big| -\tfrac1\tau\right) &= -i \alpha\, \vartheta_1(z|\tau), \quad \vartheta_{2,3,4}\left(\tfrac z\tau \big| -\tfrac1\tau\right) = \alpha\,\vartheta_{4,3,2}(z|\tau),\quad \alpha = \sqrt{-i\tau}e^{i\pi z^2/\tau}.
  \end{split}
\end{equation}
The modular $\mc S$ transformation can be proved using Poisson summation.

A useful identity relating the Dedekind eta function and the Jacobi theta functions is
\begin{equation} \label{eq:eta2theta}
  \eta^3(\tau) = \frac12 \vartheta_2(\tau) \vartheta_3(\tau) \vartheta_4(\tau), 
\end{equation}
which can be proved using the infinite product representations of the theta functions. 

The Dedekind eta is also realized as a derivative of $\vartheta_1$,
\begin{equation}\label{eq:theta1lim}
  \lim_{z\to 0}\frac1z \vartheta_1(z|\tau) = \pi \vartheta_1'(0|\tau) = 2\pi \eta(\tau)^3,
\end{equation}

The Klein-$j$ invariant can also be expressed in terms of Jacobi theta functions,
\begin{equation}\label{eq:jdef2}
    j(\tau) = 32\frac{\left(\vartheta_2(\tau)^8 + \vartheta_3(\tau)^8 + \vartheta_4(\tau)^8\right)^3}{\vartheta_2(\tau)^8 \vartheta_3(\tau)^8 \vartheta_4(\tau)^8}
\end{equation}

The following are identities involving theta functions and the Dedekind eta function when doubling and translating $\tau$ at $z=0$,
\begin{equation}
  \label{eq:thetaidentities}  
  \begin{aligned}
    \vartheta_2\left(\tfrac\tau2\right)^2 &= 2\vartheta_3\left(\tau\right)\vartheta_2\left(\tau\right), 
    & \vartheta_2\left(2\tau\right)^2 &= \frac{\vartheta_3\left(\tau\right)^2 - \vartheta_4\left(\tau\right)^2}{2}, 
    & \vartheta_2\left(\tfrac{\tau+1}{2}\right)^2 &= 2e^{\frac{i\pi}{4}}\vartheta_2(\tau) \vartheta _4(\tau),\\
    \vartheta_3\left(\tfrac\tau2\right)^2 &= \vartheta_3\left(\tau\right)^2 + \vartheta_2\left(\tau\right)^2, 
    & \vartheta_3\left(2\tau\right)^2 &= \frac{\vartheta_3\left(\tau\right)^2 + \vartheta_4\left(\tau\right)^2}{2}, 
    & \vartheta_3\left(\tfrac{\tau +1}{2}\right)^2 &= \vartheta_4(\tau)^2+i \vartheta_2(\tau)^2,\\
    \vartheta_4\left(\tfrac\tau2\right)^2 &= \vartheta_2\left(\tau\right)^2 -\vartheta_3\left(\tau\right)^2, 
    & \vartheta_4\left(2\tau\right)^2 &= \vartheta_3\left(\tau\right)\vartheta_4\left(\tau\right), 
    & \vartheta_4\left(\tfrac{\tau +1}{2}\right)^2 &= \vartheta_4(\tau)^2-i \vartheta_2(\tau)^2,\\
    \eta\left(\tfrac{\tau}{2}\right)^2 &= \vartheta_4\left(\tau\right)\eta\left(\tau\right), &
                                                                                               \eta\left(2\tau\right)^2 &= \frac12\vartheta_2\left(\tau\right)\eta\left(\tau\right), &
                                                                                                                                                                                       \eta\left(\tfrac{\tau+1}2\right)^2 &= e^{\frac{i\pi}{12}}\vartheta_3\left(\tau\right)\eta\left(\tau\right).
  \end{aligned}
\end{equation}

These can be proved by using a combination of both the product and sum definitions of the theta functions \eqref{eq:ThetaDefs}, along with their modular $\mc T$ properties \eqref{eq:TtransfTheta}.

These are useful for proving the following identity between the characters. 
First we rewrite the characters \eqref{eq:chainj} in terms of the Jacobi theta functions using \eqref{eq:jdef},
\begin{equation}
  \begin{split}
    \chi_1(\tau) &= \frac12\left(j(\tau)^{\frac23}-j(2\tau)^{\frac13}\right)
                   = \frac{\left(\vartheta_2(\tau)^8+\vartheta_3(\tau)^8+\vartheta_4(\tau)^8\right)^2}{8\eta(\tau)^{16}}
                   -\frac{\left(\vartheta_2(2 \tau)^8+\vartheta_3(2\tau)^8+\vartheta_4(2\tau)^8\right)}{4\eta(2\tau)^8},\\
    \chi_3(\tau) &= \frac12\left(j\left(\tfrac\tau2\right)^{\frac13} -(-1)^{\frac13}j\left(\tfrac{\tau+1}2\right)^{\frac13}\right) = \frac{\vartheta_2\left(\frac{\tau }{2}\right)^8+\vartheta_3\left(\frac{\tau }{2}\right)^8+\vartheta_4\left(\frac{\tau }{2}\right)^8}{4\eta \left(\frac{\tau }{2}\right)^8}
                   -\frac{(-1)^{\frac13} \left(\vartheta_2\left(\frac{\tau +1}{2}\right)^8+\vartheta_3\left(\frac{\tau +1}{2}\right)^8+\vartheta_4\left(\frac{\tau +1}{2}\right)^8\right)}{4\eta \left(\frac{\tau +1}{2}\right)^8}.
  \end{split}
\end{equation}
We use the list of identities \eqref{eq:thetaidentities}, \eqref{eq:eta2theta} and the quartic relation between the theta functions:
\begin{equation}
  \vartheta_2(\tau)^4 = \vartheta_3(\tau)^4 - \vartheta_4(\tau)^4, 
\end{equation}
to simplify both expressions, and the equality is apparent.
This can be expressed as an identity of Klein-$j$ invariants:
\begin{equation}
  \label{eq:jIdentity}
  j\left(\tfrac{\tau+1}2\right)^{\frac13} = 
  (-1)^{2/3} \left(j(\tau)^{\frac23}-j\left(\tfrac{\tau}{2}\right)^{\frac13}-j(2\tau)^{\frac13}\right).
\end{equation}

Derivatives of the theta functions can easily be derived by comparing them to the Jacobi elliptic functions $\rm sn, cn$, and $\rm dn$,
\begin{equation}
  \label{eq:el2th}
  \begin{split}
    {\rm sn}\left(\pi z\vartheta_3(\tau)^2\bigg|\frac{\vartheta_2(\tau)^4}{\vartheta_3(\tau)^4}\right)
    &=\frac{\vartheta_3(\tau)\vartheta_1(z|\tau)}{\vartheta_2(\tau)\vartheta_4(z|\tau )},\\
    {\rm cn}\left(\pi  z\vartheta_3(\tau)^2\bigg|\frac{\vartheta_2(\tau)^4}{\vartheta_3(\tau)^4}\right)
    &=\frac{\vartheta _4(\tau ) \vartheta _2(z,\tau )}{\vartheta _2(\tau ) \vartheta _4(z|\tau)},\\
    {\rm dn}\left(\pi z\vartheta_3(\tau)^2\bigg|\frac{\vartheta_2(\tau)^4}{\vartheta_3(\tau)^4}\right)
    &=\frac{\vartheta_4(\tau) \vartheta_3(z|\tau)}{\vartheta_3(\tau)\vartheta_4(z|\tau )},
  \end{split}
\end{equation}
and their derivatives
\begin{equation}
  \begin{split}
    \partial_u {\rm sn}(u|m) &= {\rm cn}(u|m)\, {\rm dn}(u|m),\\
    \partial_u {\rm cn}(u|m) &= -{\rm dn}(u|m)\, {\rm sn}(u|m),\\
    \partial_u {\rm dn}(u|m) &= -m\, {\rm cn}(u|m)\, {\rm sn}(u|m).
  \end{split}
\end{equation}
These allow us to relate derivatives of theta functions in terms of just one of the theta function derivatives,
\begin{equation} \label{eq:thetaDs}
  \begin{split}
    \vartheta_2'(z|\tau)
    &=\frac{\vartheta_2(z|\tau) \vartheta_4(z|\tau) \vartheta_1'(z|\tau)-\pi  \vartheta_2(\tau ){}^2 \vartheta_3(z|\tau) \vartheta_4(z|\tau){}^2}{\vartheta_1(z|\tau) \vartheta_4(z|\tau)},\\
    \vartheta_3'(z|\tau)
    &=\frac{\vartheta_3(z|\tau) \vartheta_4(z|\tau) \vartheta_1'(z|\tau)-\pi  \vartheta_3(\tau ){}^2 \vartheta_2(z|\tau) \vartheta_4(z|\tau){}^2}{\vartheta_1(z|\tau) \vartheta_4(z|\tau)}, \\
    \vartheta_4'(z|\tau)
    &=\frac{\vartheta_4(z|\tau) \vartheta_1'(z|\tau)-\pi  \vartheta_4(\tau ){}^2 \vartheta_2(z|\tau) \vartheta _3(z|\tau)}{\vartheta _1(z|\tau)}.
  \end{split}
\end{equation}
Higher-order derivatives can be computed by differentiating the identities above. 

The following relationships between squares of theta functions is also useful in proving many of the identities used here,
\begin{equation}\label{eq:thetaSquares}
  \begin{split}
    \vartheta_2(\tau)^2 \vartheta_1(z|\tau)^2+\vartheta_4(\tau)^2 \vartheta_3(z|\tau)^2 &=\vartheta_3(\tau)^2 \vartheta_4(z|\tau)^2,\\
    \vartheta_3(\tau)^2 \vartheta_1(z|\tau)^2+\vartheta_4(\tau)^2 \vartheta_2(z|\tau)^2 &=\vartheta_2(\tau)^2 \vartheta_4(z,\tau)^2,\\
    \vartheta_2(\tau)^2 \vartheta_2(z|\tau)^2+\vartheta_4(\tau)^2 \vartheta_4(z|\tau)^2&=\vartheta_3(\tau)^2 \vartheta_3(z|\tau)^2.   
  \end{split}
\end{equation}

\section{The WEIERSTRA{\ss} ELLIPTIC FUNCTION}
\label{sec:weierstrassellipticfunctions}

A Jacobi form of weight $k$ and index $m$ has the following modular and elliptic properties,
\begin{equation}
  \begin{split}
    \phi\left(\frac z{c\tau+d}\bigg|\frac{a\tau + b}{c\tau + d}\right) &= (c\tau+d)^k e^{\frac{2\pi i m c z^2}{c\tau+d}}\phi_k(z|\tau), \\
    \phi\left(z+\lambda \tau + \mu |\tau\right) &= e^{2\pi i m (\lambda^2 \tau + 2\lambda z)}\phi_k(z|\tau), \\
  \end{split}
\end{equation}

The Weierstra{\ss} $\wp$ function is doubly periodic and the archetypical elliptic function, defined by the series,
\begin{equation}
  \label{eq:WPfunDef}
  \begin{split}
    \wp(z;\omega_1,\omega_2) &= \frac1{z^2} + \sum_{m,n\in \mathbb Z}\left(\frac1{(z-m\omega_1 - n \omega_2)^2} - \frac1{(m\omega_1 + n \omega_2)^2}\right),\\
    \wp(z|\tau) &= \wp(z;1,\omega_2/\omega_1 = \tau).
  \end{split}
\end{equation}
It turns out that $\wp$ is a Jacobi form of weight 2 and index 0, 
\begin{equation}
  \wp\left(\frac{z}{c\tau+d}\bigg|\frac{a\tau+b}{c\tau+d}\right) = (c\tau + d)^2 \wp(z|\tau). 
\end{equation}
This can be shown by using the fact that the Laurent expansion of $\wp$ can be expressed in terms of Eisenstein series $G_{2k}$,
\begin{equation}
  \wp(z|\tau) = \frac1{z^2} + \sum_{n=1}^{\infty}(2n+1)z^{2n} G_{2n+2}(\tau), 
\end{equation}
Similarly, $\wp'$ transforms as a weight 3, index 0 Jacobi form.

Again, using the Laurent expansion, one can show that the Weierstra{\ss} $\wp$ function satisfies the following differential equation:
\begin{equation}
  \label{eq:EllipticCurve}
  \wp'(z|\tau)^2=4\wp(z|\tau)^3 -60 G_4(\tau) \wp(z|\tau) -140 G_6(\tau), 
\end{equation}
which is a result of the fact that a nonconstant elliptic function must have at least double pole or two first-order poles in the period parallelogram.

All elliptic functions can be written as rational functions of $\wp$ and $\wp'$, subject to the above differential equation \eqref{eq:EllipticCurve}.

The Weierstra{\ss} $\wp$ function and its derivative can be expressed in terms of the Jacobi theta functions,
\begin{align}\label{eq:wptotheta}
  \wp(z|\tau) &= \left(\pi  \vartheta_2(\tau) \vartheta_3(\tau) \frac{\vartheta_4(z|\tau )}{\vartheta_1(z|\tau)}\right)^2-\frac{\pi^2}{3} \left(\vartheta_2(\tau)^4+\vartheta_3(\tau)^4\right),\\
              &= \left(\pi  \vartheta_4(\tau) \vartheta_2(\tau) \frac{\vartheta_3(z|\tau )}{\vartheta_1(z|\tau)}\right)^2+\frac{\pi^2}{3} \left(\vartheta_2(\tau)^4-\vartheta_4(\tau)^4\right),\\
              &= \left(\pi  \vartheta_3(\tau) \vartheta_4(\tau) \frac{\vartheta_2(z|\tau )}{\vartheta_1(z|\tau)}\right)^2+\frac{\pi^2}{3} \left(\vartheta_3(\tau)^4+\vartheta_4(\tau)^4\right),\\                
  \wp'(z|\tau) &= -2\pi^3 \vartheta _2(\tau ){}^2 \vartheta _3(\tau )^2 \vartheta _4(\tau )^2 \frac{\vartheta_2(z|\tau )}{\vartheta_1(z|\tau)} \frac{\vartheta_3(z|\tau)}{\vartheta_1(z|\tau)} \frac{\vartheta_4(z|\tau )}{\vartheta_1(z|\tau)},
\end{align}
which can be shown with the help of \eqref{eq:thetaSquares}. 

It can be useful to express $\wp$ as an expansion in $q=e^{2\pi i \tau}$ for series solutions of differential equations,
\begin{equation}  \label{eq:WPFourier}
  \wp(z|\tau)
  = \frac{-4\pi^2}{y-2+y^{-1}} - \frac{\pi^2}{3} -4\pi^2 \sum_{n=1}^{\infty}\sum_{d|n}d(y^d - 2 + y^{-d})q^n
  = \frac{\pi^2}{\sin^2(\pi z)} - \frac{\pi^2}{3} +\pi^2 \sum_{n=1}^\infty\sum_{d|n}d\sin^2(d\pi z)q^n.
\end{equation}

\section{CORRELATOR EXPANSIONS}
\label{app:exps}
Here we list the $q$-expansions of the correlator \eqref{eq:fullcorrelator} in each unit cell of the replicated torus.
Since the correlator is two-periodic along either cycle, the correlator shows different behavior when translating one of the twist operators once around either cycle,
\begin{equation}\label{eq:exp0}
  \begin{split}
    &\ev{\sigma(z,\bar z)\sigma(0,0)}_{\tau,\bar\tau}\\
    &= \frac{\pi^2}{\left|\sin(\pi z)\right|^2}\bigg[
      1
      - \left(1-|\cos(\pi z)|^2
      -4\big(
      \sin^2(\pi z)\cos(\pi\bar z)
      +\sin^2(\pi\bar z)\cos(\pi z)
      -12 |\sin(\pi z)|^4
      \big) \right)\sqrt{q\bar q}\\
    &\quad
      + 4\sin^2(\pi z) q 
      +248\left(      
      |\cos(\pi z)|^2-\cos(\pi z)
      +4\sin^2(\pi z)(\cos(\pi\bar z)-3)\right)q\sqrt{\bar q}
    \\&\quad
    + 4\sin^2(\pi \bar z) \bar q
    +248\left(
    |\cos(\pi z)|^2-\cos(\pi \bar z)
    +4(\cos(\pi z)-3)\sin^2(\pi\bar z)\right)\bar q\sqrt q \\
    &\quad 
      +\frac12
      \Big(122994+
      123006|\cos(\pi z)|^2
      +16\big(\cos(2\pi z)+\cos(2\pi\bar z)
      -\cos(2\pi z)\cos(2\pi\bar z)\big)
    \\
    &\qquad      
      +4\big(\cos(2\pi z)\cos(\bar z) +\cos(\pi z)\cos(2\pi\bar z)\big)
      -123012\big(\cos(\pi z)+\cos(\pi\bar z)\big)
      \Big) q \bar q + \mc O(q^{\frac32}) +\mc O(\bar q^{\frac32})\bigg],
  \end{split}
\end{equation}

\begin{equation}
  \label{eq:exp1}
  \begin{split}
    &\ev{\sigma(z+1,\bar z+1)\sigma(0,0)}_{\tau,\bar\tau}\\
    &= \frac{\pi^2}{|\sin(\pi z)|^2 }\bigg[1
      -\big(1-|\cos(\pi z)|^2
      +4(\cos(\pi z)\sin^2(\pi\bar z)
      +\sin^2(\pi z)\cos(\pi\bar z)    
      -12|\sin(\pi z)|^4
      \big) \sqrt{q\bar q}\\
    &\quad
      + 4\sin^2(\pi z) q 
      +248 \big(
      |\cos(\pi z)|^2
      +\cos(\pi z)
      -4\sin^2(\pi z)(\cos(\pi\bar z)+3)
      \big)q\sqrt{\bar q}\\
    &\quad
      + 4\sin^2(\pi \bar z) \bar q
      +248 \big(
      |\cos(\pi z)|^2
      +\cos(\pi\bar z)
      -4(\cos(\pi z)+3)\sin^2(\pi\bar z)
      \big)\bar q\sqrt q\\
    &\quad
      +\frac12\Big(
      122994
      +123006|\cos(\pi z)|^2
      +16\big(\cos(2\pi z)+\cos(2\pi\bar z)-\cos(2\pi z)\cos(2\pi\bar z)\big)
    \\
    &\qquad
      -4(\cos(2\pi z)\cos(\bar z)+\cos(\pi z)\cos(2\pi\bar z))
      +123012(\cos(\pi z)+\cos(\pi\bar z))
      \Big) q\bar q + \mc O(q^{\frac32}) + \mc O(\bar q^{\frac32})
      \bigg]
  \end{split}
\end{equation}

\begin{equation}
  \label{eq:exptau}
  \begin{split}
    &\ev{\sigma(z+\tau,\bar z + \bar \tau)\sigma(0,0)}_{\tau,\bar \tau} \\
    &=\frac{\pi^2}{|\sin(\pi z)|^2} \bigg[
      3 - \cos(\pi z) \sqrt{q} - \cos(\pi\bar z)\sqrt{\bar q}
      +(|\cos(\pi z)|^2+16|\sin(\pi z)|^4 -3)\sqrt{q\bar q}\\
    &\quad
      +(12\sin^2(\pi z) - 248 \cos(\pi z))q + (12\sin^2(\pi \bar z) - 248 \cos(\pi\bar z))\bar q \\
    &\quad
      +\big(
      \cos(\pi z)+
      4(
      62\cos(\pi z)
      -\sin^2(\pi z)
      )\cos(\pi\bar z)
      -992\sin^2(\pi\bar z)
      \big)q\sqrt{\bar q}\\
    &\quad
      +\big(
      \cos(\pi\bar z)
      +4\cos(\pi z)
      (62\cos(\pi\bar z)
      -\sin^2(\pi\bar z))
      -992\sin^2(\pi z)
      \big)\bar q\sqrt q\\  
    &\quad
      +\frac12\Big(
      -615018
      +123006|\cos(\pi z)|^2
      -16(\cos(2\pi z)+\cos(2\pi\bar z) - \cos(2\pi z)\cos(2\pi\bar z))
    \\
    &\qquad\quad
      +124(\cos(\pi z)\sin^2(\pi\bar z)
      +\sin^2(\pi z)\cos(\pi\bar z))            
      \Big) q\bar q + \mc O(q^{\frac32}) + \mc O(\bar q^{\frac32})
      \bigg]
  \end{split}
\end{equation}

\begin{equation}
  \label{eq:exp1ptau}
  \begin{split}
    &\ev{\sigma(z+1+\tau,\bar z+1+\bar\tau)\sigma(0,0)}_{\tau,\bar \tau} \\
    &= \frac{\pi^2}{|\sin(\pi z)|^2} \bigg[
      3+ \cos(\pi z) \sqrt{q} + \cos(\pi\bar z)\sqrt{\bar q}
      +(|\cos(\pi z)|^2+16|\sin(\pi z)|^4 -3)\sqrt{q\bar q}
    \\
    &\quad
      +(12\sin^2(\pi z)+248\cos(\pi z))q + (12\sin^2(\pi \bar z)+248\cos(\pi\bar z))\bar q \\
    &\quad
      -\big(
      \cos(\pi z)
      -4( 62\cos(\pi z)
      +\sin^2(\pi z)
      )\cos(\pi\bar z)
      +992\sin^2(\pi\bar z)
      \big)q\sqrt{\bar q}
    \\
    &\quad
      -\big(
      \cos(\pi\bar z)
      -4\cos(\pi z)
      ( 62\cos(\pi\bar z)
      +\sin^2(\pi\bar z)
      )
      +992\sin^2(\pi z)
      \big)\bar q\sqrt q\\
    &\quad
      +\frac12\Big(
      -615018
      +123006|\cos(\pi z)|^2
      -16(\cos(2\pi z)+\cos(2\pi\bar z) - \cos(2\pi z)\cos(2\pi\bar z))
    \\
    &\qquad\quad
      -124(\cos(\pi z)\sin^2(\pi\bar z)
      +\sin^2(\pi z)\cos(\pi\bar z))            
      \Big) q\bar q + \mc O(q^{\frac32}) + \mc O(\bar q^{\frac32})
      \bigg]
  \end{split}
\end{equation}

\newpage
\section{FIGURES}
\label{app:figures}
\begin{figure}[!htbp]
  \centering
  \begin{subfigure}[t]{0.45\textwidth}
    \centering
    \includegraphics[height=45mm]{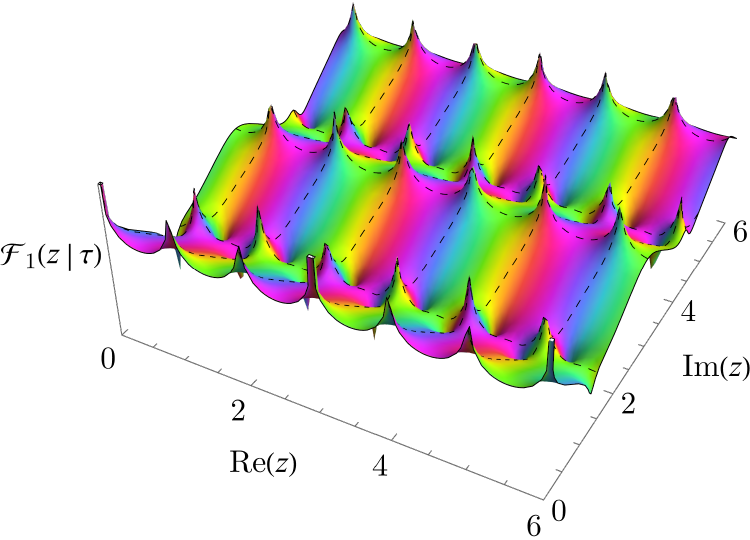}
    \caption*{}
  \end{subfigure} 
  \begin{subfigure}[t]{0.45\textwidth}
    \centering
    \includegraphics[height=45mm]{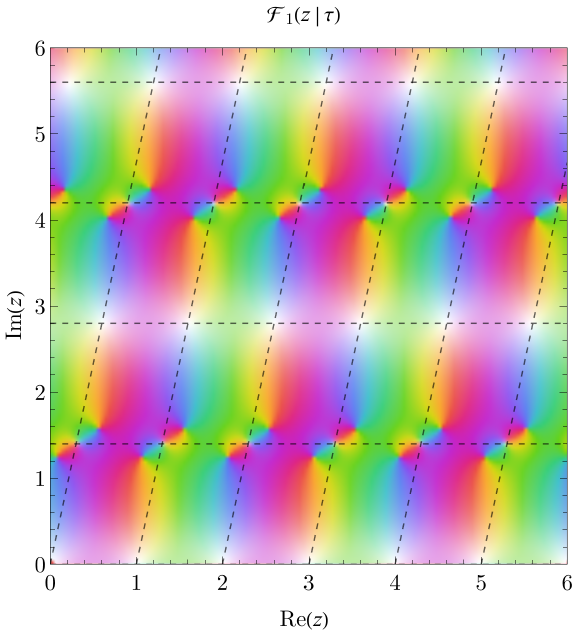}
    \caption*{}
  \end{subfigure}
  \begin{subfigure}[t]{0.45\textwidth}
    \centering
    \includegraphics[height=45mm]{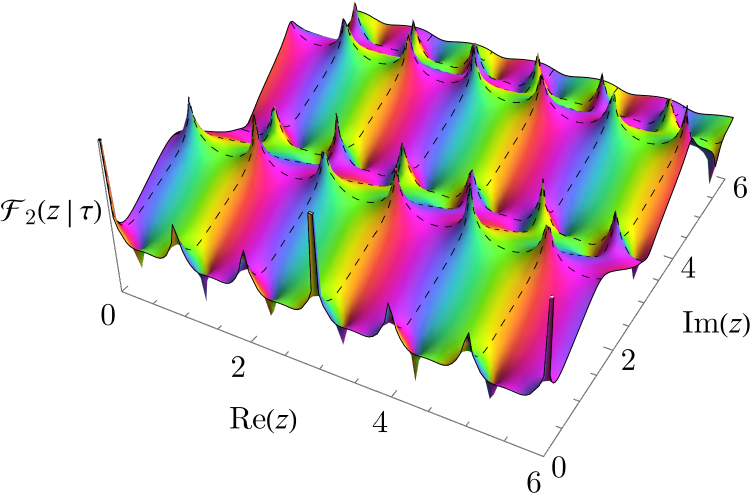}
    \caption*{}
  \end{subfigure} 
  \begin{subfigure}[t]{0.45\textwidth}
    \centering
    \includegraphics[height=45mm]{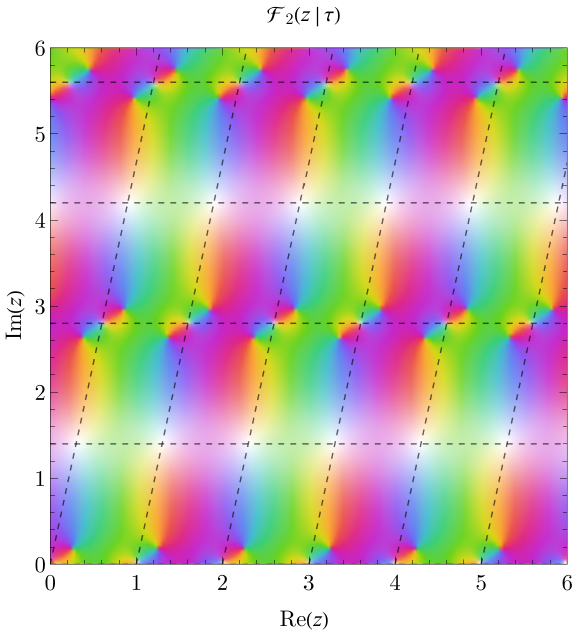}
    \caption*{}
  \end{subfigure}
  \begin{subfigure}[t]{0.45\textwidth}
    \centering
    \includegraphics[height=45mm]{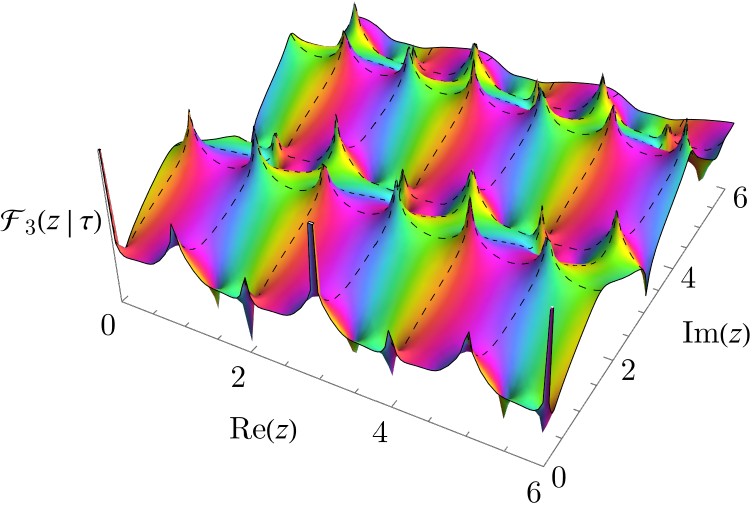}
    \caption{}
  \end{subfigure} 
  \begin{subfigure}[t]{0.45\textwidth}
    \centering
    \includegraphics[height=45mm]{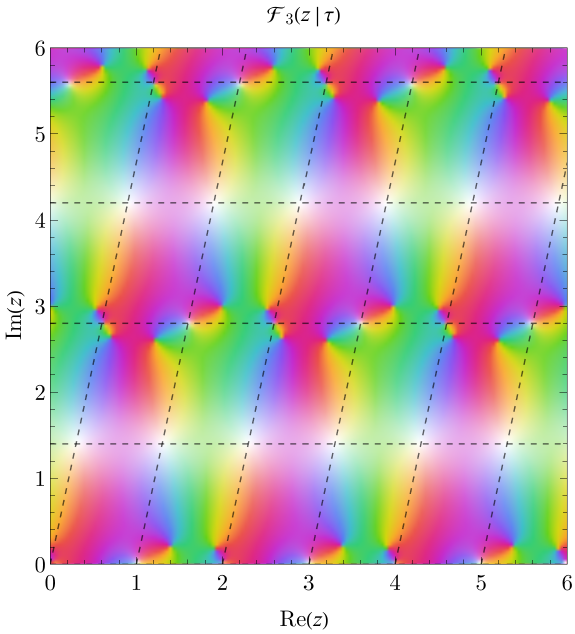}
    \caption{}
  \end{subfigure}
  \begin{subfigure}[t]{\textwidth}
    \centering
    \includegraphics[width=55mm]{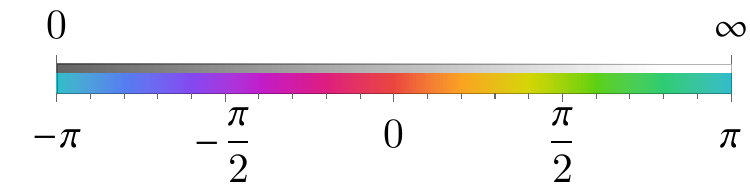}
    \caption*{}
  \end{subfigure}
  \caption{\justifying
    Figures of the conformal blocks \eqref{eq:confblocks} are presented for the value $\tau = 0.3 + 1.4 i$ within the range extending from 0 to $6+6i$.
    In the left panel (a), a three-dimensional visualization of the conformal blocks is illustrated. 
    Here, the height (set on a logarithmic scale) represents the magnitude of the block $\mc F_i(z|\tau)$, with phase information illustrated through a color gradient (using a rainbow scale, as indicated in the legend). 
    Meanwhile, the right panel (b) provides a two-dimensional visualization of the conformal blocks $\mc F_i(z|\tau)$ where the opacity reflects the magnitude, and the poles are specifically highlighted in white.
    In both diagrams, dashed lines indicate the toroidal lattice configurations determined by the vectors 1 and $\tau$.}
  \label{fig:conformalblocks}
\end{figure}

\twocolumngrid

\bibliography{torbib}

\end{document}